\documentclass[fleqn,usenatbib,useAMS]{mnras}

\usepackage[switch,mathlines]{lineno}

\usepackage{graphicx}
\usepackage{amsmath}
\usepackage{mathtools}
\usepackage{booktabs}
\usepackage{longtable}
\usepackage{multirow}
\usepackage[normalem]{ulem}
\usepackage{xcolor}
\usepackage{comment}
\usepackage{changepage}
\usepackage{tabularray}
\graphicspath{{./}{Figure/}}

\usepackage{orcidlink}

\newcommand{\NC}{\textsc{NewCluster}}

\newcommand{\SKIRT}{\textsc{SKIRT}}

\definecolor{Green}{rgb}{0.1,0.7,0.2}

\newcommand{\accup}{\textcolor{green}{$\uparrow$}}
\newcommand{\accdown}{\textcolor{red}{$\downarrow$}}

\newcommand\edel{\bgroup\markoverwith{\textcolor{red}{\rule[0.5ex]{2pt}{0.8pt}}}\ULon}

\usepackage[T1]{fontenc}
\usepackage{ae,aecompl}
\usepackage{newtxtext,newtxmath}

\title[High-z dust growth in NewCluster spin-off]{Dust growth in the early Universe: insight from the NewCluster spin-off simulation}

\author[J. K. Jang et al.]{
J. K. Jang\,\orcidlink{0000-0002-0858-5264},$^{1}$\thanks{E-mail: \href{mailto:starbrown816@gmail.com}{starbrown816@gmail.com} (JKJ); \href{mailto:yi@yonsei.ac.kr}{yi@yonsei.ac.kr} (SKY)}
Yohan Dubois\,\orcidlink{0000-0003-0225-6387},$^{2}$
Francisco Rodr\'{\i}guez Montero\,\orcidlink{0000-0001-6535-1766},$^{3}$
San Han\,\orcidlink{0000-0001-9939-713X},$^{2}$
Jinsu Rhee\,\orcidlink{0000-0002-0184-9589},$^{2}$
\newauthor
Diana Ismail\,\orcidlink{0009-0007-2281-4944},$^{4}$
Gyeong-Hwan Byun\,\orcidlink{0009-0007-8611-3813},$^{1}$
Changyeon Yu\,\orcidlink{0009-0007-9124-894X},$^{1}$
Seyoung Jeon\,\orcidlink{0000-0002-1270-4465},$^{5}$
and Sukyoung K. Yi\,\orcidlink{0000-0002-4556-2619} $^{1}$
\\
$^{1}$Department of Astronomy, Yonsei University, Seoul 03722, Republic of Korea\\
$^{2}$Institut d'Astrophysique de Paris, Sorbonne Universit\'{e}, CNRS, UMR 7095, 98 bis bd Arago, 75014 Paris, France\\
$^{3}$Kavli Institute for Cosmological Physics, University of Chicago, Chicago, IL 60637, USA\\
$^{4}$Observatoire Astronomique de Strasbourg, Universit\'{e} de Strasbourg, CNRS, UMR 7550, 67000 Strasbourg, France\\
$^{5}$Korea Astronomy and Space Science Institute, 776 Daedeok-daero, Yuseong-gu, Daejeon 34055, Republic of Korea
}

\date{Accepted 2026 October 28. Received 2026 October 13; in original form 2026 June 19}

\pubyear{\the\year{}}

\begin{document}
\label{firstpage}
\pagerange{\pageref{firstpage}--\pageref{lastpage}}
\maketitle

\begin{abstract}
We investigate dust evolution in galaxies at $z\sim5$ using a spin-off run of the \NC\ cosmological zoom-in simulation with a live two-species, two-size dust model. 
We find that dust evolution follows three metallicity-dependent phases. 
At $\log(Z/Z_\odot)<-1$, dust is dominated by stellar ejecta and produces a shallow, featureless UV extinction curve whose normalized wavelength dependence is similar to that of the Calzetti attenuation law (Phase I).
At $-1\leq \log(Z/Z_\odot)<-0.5$, efficient silicate accretion drives rapid dust growth, producing silicate-rich dust and SMC-like extinction curves (Phase II). 
At $\log(Z/Z_\odot)\geq-0.5$, carbonaceous grain growth also becomes efficient, yielding a more balanced grain mixture and Milky-Way-like extinction curves (Phase III). 
In addition, local dust-to-metal ratios are strongly suppressed in regions affected by recent clustered supernova feedback, further complicating the interpretation of high-$z$ observations.
Despite some successes, our galaxies lie below current observationally inferred dust masses and dust-to-metal ratios for high-redshift galaxies.
The resulting depletion patterns suggest that the fixed Fe-bearing olivine composition assumed in our current dust model may be too restrictive for chemically young galaxies, and that metallicity-dependent grain compositions may be required to explain the dust content of the early Universe.

\end{abstract}

\begin{keywords}
galaxies: evolution -- galaxies: high-redshift -- galaxies: ISM -- dust, extinction -- methods: numerical
\end{keywords}

\section{Introduction}
\label{sec:introduction}

The study of cosmic dust has progressed from simple extinction measurements to a mature framework that links grain formation pathways, interstellar processing, and multi-wavelength observables.
In the high-redshift Universe, where galaxy assembly, chemical enrichment, and stellar feedback proceed on short dynamical timescales, dust is not merely a passive opacity source but an active regulator of cooling, chemistry, and the spectral energy distribution of galaxies.
Modern dust-evolution theory took shape when stellar dust production was first coupled to galactic chemical evolution \citep{Dwek&Scalo80,Dwek98}.
The interpretation of the unidentified infrared emission bands in terms of polycyclic aromatic hydrocarbons (PAHs) \citep{Leger&Puget84}, together with models of grain-size distributions and dust extinction and emission \citep{MRN77,Desert90,Weingartner&Draine01}, highlighted the importance of grain composition and size distribution in shaping dust observables.
By the 2000s--2010s, gas-phase metal accretion in dense interstellar medium (ISM) phases had emerged as a key, and often dominant, channel for rapid dust growth \citep{Edmunds01,Asano13,Hirashita15}.

Observationally, the Infrared Astronomical Satellite (IRAS) established the importance of dust emission in the luminosity budget of the local Universe \citep{Soifer87}, with the Spitzer Space Telescope subsequently enabling more sensitive infrared imaging and spectroscopy \citep{Werner04}.
The discovery of substantial dust masses at $z>5$ \citep{Bertoldi03} challenged simple expectations that young galaxies tend to be dust-poor, and motivated dust models that can operate under rapidly evolving and chemically immature conditions.
More recently, the Reionization Era Bright Emission Line Survey (REBELS; \citealt{Bouwens22}) and the ALMA Large Program to Investigate C$^+$ at Early Times (ALPINE; \citealt{Faisst20,Bethermin20}) have provided systematic measurements of dust-continuum and ISM emission in galaxies at $z\sim5$--$7$. 
Follow-up analyses have inferred dust masses and temperatures for these samples \citep{Ferrara22,Pozzi21,Sommovigo22,Algera26}, while JWST provides complementary constraints on attenuation curves and stellar populations \citep{Markov23,Woodrum25,Shivaei26}.
Together, these studies reveal substantial galaxy-to-galaxy diversity in inferred dust properties and motivate models that follow dust formation and processing in rapidly evolving high-redshift environments.

On the theoretical side, analytic and semi-analytic models have provided much of the foundational understanding of dust production, growth, and destruction \citep{Zhukovska08,Hirashita15}.
At the same time, dust models connecting carbonaceous and silicate grain populations to extinction and emission observables, such as the framework of \citet{Weingartner&Draine01} and the heterogeneous dust evolution model for interstellar solids (THEMIS; \citealt{Jones17}), have clarified how grain composition and size distribution shape observable dust signatures.
Recent implementations of live dust evolution enable studies of dust formation and processing in hydrodynamic simulations, including SIMBA \citep{Dave19,Li19}, SIMBA-EoR \citep{Jones24}, FIRE-2 \citep{Hopkins18,Choban22,Choban25}, COLIBRE \citep{Schaye26,Trayford26}, Cosmic Sands \citep{Lower23}, and simulations using AREPO with SMUGGLE \citep{Narayanan23,Narayanan25,Narayanan26}.

Despite these advances, high-redshift galaxies remain a stringent test of current dust prescriptions.
Their short enrichment timescales, bursty star-formation histories, strong clustered feedback, and non-solar abundance patterns can amplify uncertainties that are less apparent in local, Milky-Way-like calibration environments.
Key unresolved issues include condensation efficiencies and reverse-shock survival \citep{Kirchschlager23}, size-dependent shattering and coagulation \citep{Yan04,Moseley23}, dust--gas dynamical coupling \citep{Commercon23}, thermal sputtering in hot gas \citep{Bocchio14,Slavin15}, and the phase-dependent processing of dust in a turbulent, multiphase ISM \citep{Kirchschlager24}.
In addition, the observed diversity of extinction curves \citep{Gordon03,Fitzpatrick&Massa07} continues to stress-test assumptions about grain sizes, compositions, and processing histories.
These uncertainties make forward modelling and mock surveys essential for connecting theoretical dust models to high-$z$ observations while accounting for selection effects and measurement systematics \citep{Hayward11,Trayford17,Byun25}.

Against this backdrop, we analyse dust formation and evolution in high-$z$ galaxies using a spin-off run of the \NC\ cosmological zoom-in simulation that implements the live dust model of \citet{Dubois24}.
This simulation follows two grain species and two grain-size bins, together with stellar dust production, gas-phase metal accretion, shattering, coagulation, thermal sputtering, and supernova-driven destruction, within a fully cosmological environment.
Our objective is twofold.
First, we examine how this two-species, two-size, multi-process dust prescription behaves in galaxies at $z\sim5$ and how its predictions compare with current high-$z$ dust constraints.
Second, we identify which assumptions in the model---especially those related to grain composition, sub-grid accretion physics, and feedback-driven destruction---limit agreement with the data.
In doing so, we aim to isolate the physical processes that dominate early dust evolution and to provide concrete directions for the next generation of cosmological dust models.


\begin{figure*}
\centering
\includegraphics[width=0.95\textwidth]{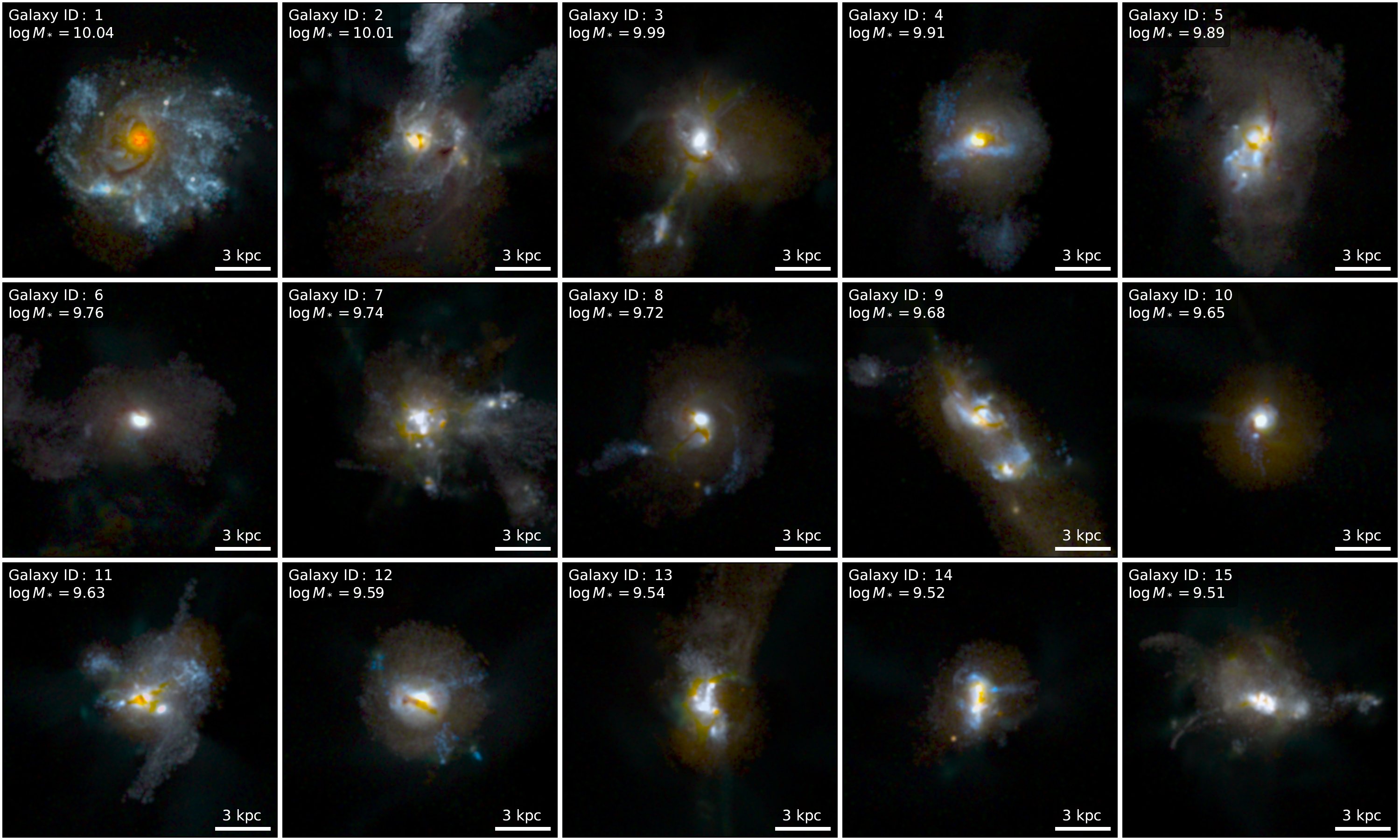}
\caption{
Example mock images of galaxies at redshift $z=5$ from the \NC\ spin-off simulation (NCso).
All images are generated using the live dust output of NCso and post-processed with the radiative transfer code \SKIRT.
The JWST F070W, F115W, and F200W filters are assigned to the blue, green, and red channels, respectively.
For visual purposes, neither instrumental noise nor point spread function (PSF) convolution is included.
}
\label{fig:Figure1}
\end{figure*}

\section{Methodology}

\subsection{Simulation}
\label{sec:sub-simulation}
We utilize the \NC\ spin-off simulation (hereafter NCso) for our analysis.
The NCso simulation was evolved down to $z \sim 5$, which sets the lowest redshift available in this work.
NCso is essentially identical to the \NC\ simulation (hereafter NC), except for the adopted dust prescription.
NC is a zoom-in cosmological hydrodynamic simulation targeting a galaxy cluster with a virial mass of
$4.7 \times 10^{14}\, M_\odot$ at $z=0$.
The detailed subgrid physics implemented in the \NC-series simulations is presented in \citet{Han25}.
Here, we briefly summarize the subgrid physics common to both simulations.

The simulations are performed with RAMSES-yOMP \citep{Han25yomp}, a specific branch of the adaptive mesh refinement (AMR) code RAMSES \citep{Teyssier02}. 
RAMSES-yOMP utilizes both the message passing interface (MPI) and open multiprocessing (OpenMP) libraries for parallel computation, thereby improving computational speed and load balancing.
The initial conditions of the zoom-in region are generated using MUSIC \citep{Hahn&Abell11}.
We adopt the WMAP7 \citep{Komatsu11} cosmology with parameters
$\Omega_b = 0.0455$, $\Omega_m = 0.272$, $\Omega_\Lambda = 0.728$, $h_0 = 0.703$,
$n_{\rm s} = 0.967$, and $\sigma_8 = 0.810$ for constructing the initial conditions within a total volume of
$(100\,h^{-1}\,{\rm Mpc})^3$.
To minimize contamination from outside the zoom-in region, we determine the region through the following procedure.
We first run a dark matter--only (DM-only) simulation and identify all DM particles within 3.5 virial radii (3.5 $R_{\rm vir}$) of the target halo at $z=0$.
These particles are then traced back to their initial positions at $z=50$.
We define the convex-hull region that encloses all of these particles, which sets the boundary of the final zoom-in region.
The DM mass resolution is set to level 12, corresponding to a particle mass of
$1.3 \times 10^{6}\, M_\odot$.
We allow refinement up to level 21, yielding a spatial resolution of 68 pc at $z=0$.
The minimum stellar particle mass is $\sim 2.5 \times 10^{4}\,M_\odot$.
To reduce potential numerical shot noise in the gravitational calculation arising from the large mass ratio between DM and stellar particles, we cap the maximum refinement level used for the DM gravitational potential at level 18, while stellar particles are refined up to level 21.

Stellar feedback includes Type II and Type Ia supernovae, as well as stellar winds.
Feedback events deposit momentum and energy through the mechanical scheme of \citet{Kimm&Cen14}, with a density-dependent boost factor that ensures the correct momentum injection during the unresolved Sedov--Taylor phase \citep{Kimm17}.
SN-II explosions occur once a star particle reaches an age of $5\,\rm Myr$, with a specific energy budget of $0.045 \times 10^{51}\,{\rm erg}\ M_\odot^{-1}$.
SN-Ia events follow a power-law delay-time distribution, $\Psi(t)\propto t^{-1}$ \citep{Maoz12}, normalized such that a simple stellar population (SSP) of mass $M_{\rm SSP}$ produces $N_{\rm SNIa}/M_{\rm SSP}=0.0013\,M_\odot^{-1}$ over $0.05$--$13.7\,\rm Gyr$.
Each SN-Ia releases $10^{51}\,\rm erg$.

The radiative heating and cooling processes follow the framework of \citet{Dubois21}.
A spatially uniform UV background from \citet{Haardt&Madau96} is adopted, with the re-ionization redshift set to $z_{\rm reion}=10$.
To approximate self-shielding in dense gas, we reduce the UV photo-heating rate by the exponential factor adopted by \citet{Dubois21}, motivated by the radiation-hydrodynamic results of \citet{Rosdahl&Blaizot12}: $\exp[-n_{\rm H}/(0.01\,{\rm H\,cm^{-3}})]$. 
This prescription strongly suppresses UV photo-heating at $n_{\rm H}\gtrsim0.01\,{\rm H\,cm^{-3}}$.
Metallicity-dependent cooling is computed using the tabulated rates of \citet{Sutherland&Dopita93} and \citet{Dalgarno&McCray72}, where metallicity is defined as the total abundance of all gas-phase elements except H, He, and D.
At high temperatures, the simulation also incorporates cooling from electron--dust collisional interactions.
We adopt the rates of \citet{Dwek&Werner81}, which depend on the gas temperature, grain size, and the number densities of hydrogen and electrons, assuming a fully ionized medium.
This dust-related cooling becomes negligible below $\sim 10^6\ \mathrm{K}$, because of the lower rate of the electron-grain collisions.

For chemical enrichment, we account for mass return from stellar winds and from both type-II and Ia supernovae.
A \citet{Chabrier03} initial mass function is adopted throughout.
Each stellar particle is treated as a simple stellar population, for which Starburst99 \citep{Leitherer99,Leitherer14} provides the time-dependent mass-return history, stellar-wind feedback, and supernova rates.
We track the enrichment of ten species (H, D, He, C, N, O, Mg, Fe, Si, and S).
The elemental yields of SN-II ejecta are then assigned using the tables of \citet{Kobayashi06}, assuming progenitor masses of $8$--$50\,M_\odot$.
The contribution from stellar winds follows the Geneva stellar-evolution models \citep{Schaller92,Maeder&Meynet00}, with the associated energy injected locally into the grid cell hosting the star particle.
For SN-Ia, we adopt the yields of \citet{Iwamoto99}, with each explosion producing an ejecta mass near the Chandrasekhar limit ($\sim 1.4\,M_\odot$).
Together, these processes ensure a self-consistent evolution of elemental abundances in the simulation.
The more detailed prescription for NC can be found in \citet{Han25} and \citet{Byun25}.

\subsection{Dust modelling}
\label{sec:sub-dust model}
Here, we describe the dust model used in the NCso simulation.
The overall dust treatment in the simulation follows the framework of \citet{Dubois24}.
The purpose of this spin-off run is to implement and test the dust evolution model of \citet{Dubois24} in the high-$z$ environment.
Below, we summarize the dust modelling adopted in our simulation.

\subsubsection{Grain species}
\label{sec:subsub-creation}
We assume two distinct dust grain populations in our model: amorphous carbonaceous grains and silicate grains with a fixed olivine composition.
The silicate grains are modelled exclusively as spherical olivine, $\rm Mg_{2-x}Fe_xSiO_4$, with $x=1$.
Based on the analysis of \citet{Min07}, which shows that the inferred extinction of interstellar dust grains of MW is well explained by the spherical grains composed of amorphous olivine and pyroxene---with more than 50\% of the silicate mass in olivine with $x=1$---we adopt this composition to approximate the bulk MW dust properties, acknowledging that the full complexity of dust mineralogy remains beyond the scope of this study.
We assume solid carbonaceous grains with a material density of $s_{\rm C}=2.2\,{\rm g\,cm^{-3}}$ and silicate grains with $s_{\rm Sil}=3.3\,{\rm g\,cm^{-3}}$.
We implement the two-size approximation following \citet{Hirashita15TwoSize}. 
Following the implementation of \citet{Dubois24}, we adopt characteristic radii of $5\,{\rm nm}$ for the small grain bin and $0.1\,{\rm \mu m}$ for the large grain bin.
Although such a two-size approximation inevitably omits fine details of the grain size distribution, it captures the bulk behaviour of dust while avoiding the substantial computational cost associated with fully resolved size distributions---both in memory usage and runtime.

\subsubsection{Dust formation and growth}
\label{sec:subsub-formation}
Dust grains are produced and grow through two pathways:
(1) the formation of grain cores in stellar ejecta from SN-II, SN-Ia, and AGB stellar winds, and
(2) the subsequent accretion of gas-phase metals onto the surfaces of pre-existing grains.

The amount of dust formed in stellar ejecta is determined by the condensation efficiency, $\delta$.
For silicate grains, we follow the implementation of \citet{Dubois24} and adopt condensation efficiencies of $\delta_{\rm Sil}^{\rm SNII}=0.8$ and $\delta_{\rm Sil}^{\rm SNIa}=1$. 
The SN-II value follows \citet{Dwek98}.
The amount of silicate dust produced by each stellar particle is determined by the limiting element, defined as the element that is least abundant relative to the adopted silicate composition.
For carbonaceous grains, we assume $\delta_{\rm C}^{\rm SNII}=\delta_{\rm C}^{\rm SNIa}=0.5$.
Regarding AGB stellar winds, the condensation efficiency depends on the carbon-to-oxygen ratio (C/O).
If $\mathrm{C/O} < 1$, oxygen remains available to form silicate dust, giving $\delta_{\rm Sil}^{\rm AGB,C/O<1}=0.8$, while carbonaceous dust does not form ($\delta_{\rm C}^{\rm AGB,C/O<1}=0$).
If $\mathrm{C/O} \geq 1$, all available oxygen is locked into CO molecules, preventing silicate formation ($\delta_{\rm Sil}^{\rm AGB,C/O\geq1}=0$), while carbonaceous grains form efficiently ($\delta_{\rm C}^{\rm AGB,C/O\geq1}=1$).
For computational simplicity, we assume that all dust grains released by stellar ejecta belong to the large grain population.

Gas-phase metal accretion onto the surfaces of pre-existing grain cores is included following the classical formulation of \citet{Dwek98}.
For each dust species $i$ and size bin $j$, the accretion rate is given by
\begin{equation}
    \dot{M}_{{\rm acc},i,j} = \left(1-\frac{M_{{\rm D},i,j}}{M_{{\rm Z},i}}\right) \frac{M_{{\rm D},i,j}}{t_{{\rm acc},i,j}},
\end{equation}
where $M_{D,i,j}$ is the local dust mass, $M_{Z,i}$ is the mass of the limiting element (gas+dust), and $t_{acc,i,j}$ is the accretion timescale.
For silicates, the limiting element is the least abundant among Mg, Fe, Si, and O, while for carbonaceous grains it is C.
The accretion timescale is
\begin{equation}
\begin{split}
t_{{\rm acc},i,j} =\;& 0.28\,\alpha^{-1}\,a_{{\rm 0.005},j}\,s_{{\rm 3},i} \\
&\times \frac{A^{1/2}_{\rm X} f_{\rm X}}{E_i(a_j) Z_{\rm X}}
\left(\frac{1\,{\rm H\,cm^{-3}}}{n}\right)
\left(\frac{50\,{\rm K}}{T}\right)^{1/2}
\,{\rm Myr},
\end{split}
\end{equation}
where $\alpha$ is the sticking coefficient, $a_{{\rm 0.005},j}=a_{j}/5\, {\rm nm}$, $s_{{\rm 3},i}=s_{i}/3\, {\rm g\, cm^{-3}}$, $f_{\rm X}$ and $A_{\rm X}$ are the mass fraction in the composition of grain and the atomic weight of the limiting element, respectively, $E_{i}(a_j)$ is the Coulomb enhancement factor, $Z_{\rm X}$ is the mass fraction of the limiting element in total (gas+dust), and $n\ {\rm and}\ T$ are the density and temperature of the gaseous cell.
The Coulomb factor $E_i(a_j)$ accounts for electrostatic enhancement or suppression of collisions between gas-phase ions and dust grains \citep{Weingartner&Draine99}.
Following \citet{Dubois24}, we adopt fixed representative factors motivated by the CNM curves in Fig.~1 of \citet{Weingartner&Draine99}.
We set $E=1$ for all grains, except in cells with $n>10\,{\rm H\,cm^{-3}}$ and $T<2\times10^4\,{\rm K}$, where we adopt $E_{\rm Sil}(5\,{\rm nm})=10$ and $E_{\rm C}(0.1\,{\rm \mu m})=0$.
The latter choice switches off accretion onto large carbonaceous grains in these cells.
These fixed values provide a coarse approximation for our two-size model: the reference curves assume specific gas conditions and a radiation field, whereas we do not recompute the factors from the local ionization state or grain-charge distribution.
To account for accretion occurring within unresolved cells due to the limited spatial resolution of the simulation, we apply a turbulence-driven sub-grid correction.
The gas density within a cell of mean value $\bar{n}$ is assumed to follow a log-normal distribution with a variance of $\sigma^2_s={\rm ln}(1+b^2 \mathcal{M}^2)$, where $b=0.4$ and $\mathcal{M}$ is the Mach number.
Integrating the accretion rate over this PDF yields an effective timescale that accelerates accretion in the high-density and low-temperature gas.
This correction is applied only to the gaseous cell with $\bar{n}>0.1\, {\rm H\, cm^{-3}}$, $T<10^4\ {\rm K}$, and a Jeans length unresolved at the grid scale.
Following \citet{Dubois24}, we adopt a constant effective sticking coefficient of $\alpha_{\rm eff}=1/3$ for the unresolved accretion regime described above.
For carbonaceous grains, we assume that gas-phase carbon is fully locked into CO above an adopted threshold $n_{\rm thr}=5\times10^3\, {\rm H\,cm^{-3}}$ in NCso.
This threshold is higher than the value adopted in the calibration run of \citet{Dubois24} ($n_{\rm thr}=1\times10^3\,{\rm H\,cm^{-3}}$) and is adopted phenomenologically to represent the possibility of elevated cosmic-ray ionization rates in actively star-forming high-redshift galaxies.
Because the cosmic-ray ionization rate is not evolved explicitly in NCso, this threshold should be regarded as a model assumption rather than a self-consistently calculated CO transition density.

\begin{figure*}
\centering
\includegraphics[width=0.95\textwidth]{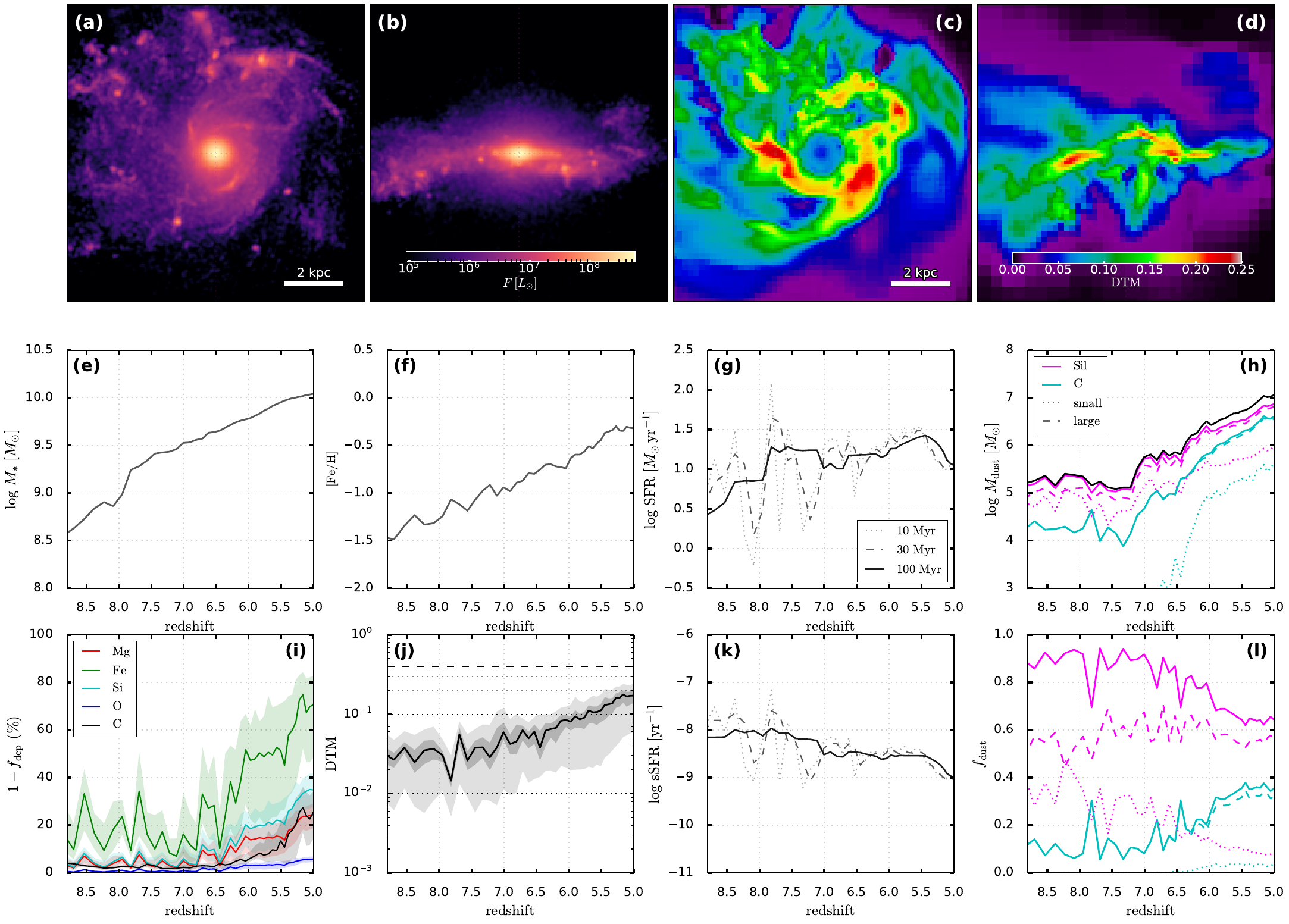}
\caption{
Example galaxy properties from the \NC\ spin-off simulation (NCso).
The first-row panels show face-on and edge-on projections of the galaxy at the final snapshot ($z\sim5$).
The left two panels display the projected SDSS $g$-band luminosity, while the right two panels show the projected dust-to-metal mass ratio (DTM).
The remaining panels illustrate the evolution of key physical properties measured within three half-mass radii ($3R_{\rm 50}$).
Shown are the stellar mass, metallicity ([Fe/H]), star-formation rate (SFR) and specific SFR (sSFR; each measured using three different time windows), depletion fractions of dust-forming elements, the global DTM, dust masses of individual grain species (magenta and cyan denote silicate and carbonaceous grains, respectively, with dotted and dashed lines indicating the small- and large-grain bins), and the evolution of small- and large-grain fractions from redshift 9 to 5.
}
\label{fig:Figure2}
\end{figure*}

\begin{figure*}
\centering
\includegraphics[width=0.95\textwidth]{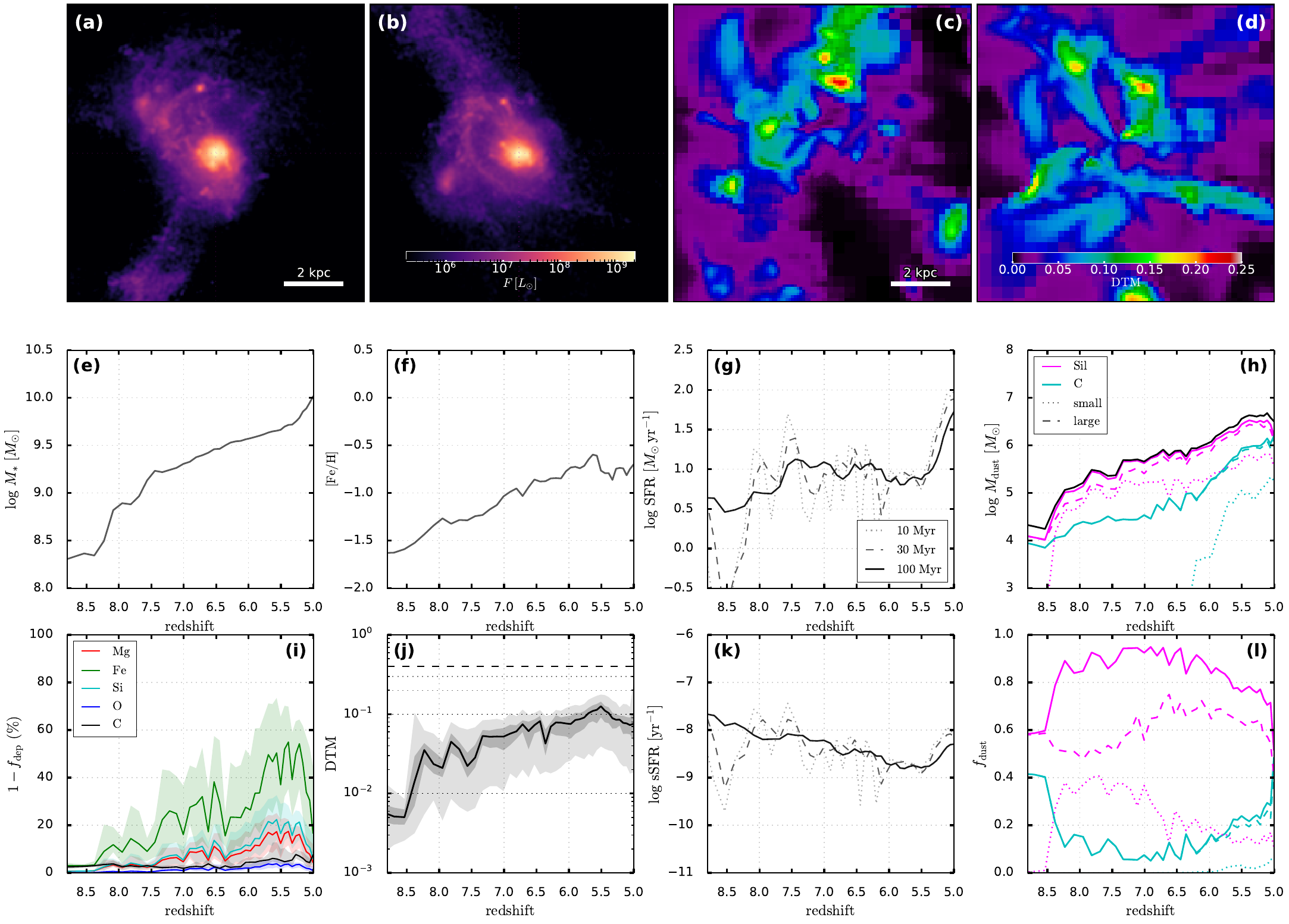}
\caption{
Same as Figure~\ref{fig:Figure2}, but for a merger-affected galaxy that undergoes a recent starburst near $z\sim5.25$.
In contrast to the disk-forming system in Figure~\ref{fig:Figure2}, this galaxy shows a more disturbed gas/dust structure and a stronger suppression of the dust content following clustered SN-II feedback.
}
\label{fig:Figure3}
\end{figure*}

\subsubsection{Dust Destruction}
\label{sec:subsub-destruction}
Dust destruction in our model includes both thermal sputtering and supernova-driven destruction.

Thermal sputtering provides a destruction channel whenever grains reside in gas hotter than $T\sim10^6\ {\rm K}$.
Using the fitting formula of \citet{Hu19}, the sputtering timescale is given by
\begin{equation}
    t_{{\rm spu},i,j} = \frac{a_{i,j}}{3\, n_{\rm H}\, Y_{{\rm th},i}(T)},
\end{equation}
where $a_{i,j}$ is the grain size, $n_{\rm H}$ is the hydrogen number density of the gas, and $Y_{{\rm th},i}(T)$ is the thermal sputtering yields calculated by \citet{Nozawa06}. 
We note that sputtering yields for carbonaceous grains are significantly lower (by a factor of $\sim 3$) at high temperatures.
Although sputtering has been discussed as a potentially strong destruction mechanism in hot halos, \citet{Dubois24} find it to be subdominant in idealized disk simulations compared to direct SN-driven destruction: even artificially increasing the sputtering timescale by a factor of ten leads to negligible changes in the total dust mass in the ISM.

The destruction induced by the supernovae is considered as a combination of the thermal \& non-thermal sputtering and the grain collisions \citep{Kirchschlager22}. 
However, because our simulation lacks the detailed information required to compute these processes explicitly, SN-driven destruction is included through a sub-grid prescription.
For type-II and Ia supernovae, we adopt the size-dependent destruction formalism of \citet{Aoyama20}, in which the fraction of dust removed per event depends on $M_{100}$, the shocked gas mass above $100\, \rm km\, s^{-1}$:
\begin{equation}
    \Delta M_{{\rm SN},i,j}=\epsilon_{{\rm SN},i}(a_j)\,{\rm min}\left(\frac{M_{100}}{M_{\rm g}},1\right)\, M_{{\rm D},i,j},
\end{equation}
where $M_{\rm g}$ is the mass of the gas, $M_{{\rm D},i,j}$ is the local dust mass, and the size-dependent destruction efficiency $\epsilon_{{\rm SN},i}(a_j)$ given by
\begin{equation}
    \epsilon_{{\rm SN},i}(a_j)=1-{\rm exp}\left(-\frac{\delta_{{\rm SN},i}}{a_{0.1,j}}\right),
\end{equation}
where $\delta_{{\rm SN},i}$ is the SN destruction efficiency and $a_{0.1,j}$ is the dust grain size normalized by 0.1 $\mu \rm m$.
Small grains suffer more severe destruction because drag forces confine them within the hot post-shock region, where sputtering proceeds more rapidly.
When multiple supernovae occur within a single timestep, the net mass loss is accumulated using the recursive formulation of \citet{Hou17}.
We adopt different efficiencies for carbonaceous and silicate grains, $\delta_{\rm SN,C}=0.10$ and $\delta_{\rm SN,Sil}=0.15$, to be consistent with the behaviour that silicates are more fragile in SN shocks \citep{Hu19}.

\subsubsection{Size Evolution}
\label{sec:subsub-size}

The redistribution of mass between the small and large grain populations is governed by two processes: shattering in diffuse, turbulent gas and coagulation in dense, cold environments.
Both mechanisms conserve the total dust mass while altering the grain-size distribution.

Shattering converts large grains into small ones when grain--grain collisions occur with sufficiently high relative velocities.
The corresponding timescale is set by the grain collision rate and scales with grain size, density, and large-grain velocity dispersion.
Following \citet{Granato21}, we adopt the simplified form:
\begin{equation}
    t_{\rm sha}=54\, a_{0.1}\, s_3\, \left(\frac{0.01}{D_{\rm L}}\right)\, \left(\frac{n_{\rm H}}{1\, {\rm H\, cm^{-3}}}\right)^{-p_{\rm sh}}\ {\rm Myr},
\end{equation}
where $a_{0.1}$ is the radius of the large grain normalized by $0.1\,\mu{\rm m}$, $s_3$ is the grain material density normalized by $3\, {\rm g\,cm^{-3}}$, $D_{\rm L}$ is the dust-to-gas ratio of large grains, and $n_{\rm H}$ is the hydrogen number density.
The exponent varies with density to mimic the decrease in turbulent grain velocities from the warm ionized medium to molecular-cloud densities: $p_{\rm sh}=1$ for $n_{\rm H}<1\, {\rm H\,cm^{-3}}$ and $p_{\rm sh}=1/3$ for $1\leq n_{\rm H}<10^3\, {\rm H\,cm^{-3}}$.
The shattering prescription is evaluated only over $n_{\rm H}<10^3\, {\rm H\,cm^{-3}}$; at higher densities, collisional grain processing is represented by coagulation.
Since the threshold velocity for fragmentation is a few $\rm km\, s^{-1}$, shattering is primarily active in the diffuse ISM where large grains reach the turbulent forcing-scale velocities.

In dense, cold gas, the relative velocities of small grains drop below the coagulation threshold ($\sim 0.1-1\, {\rm km\, s^{-1}}$), enabling collisions that merge small grains into larger ones.
We model the coagulation timescale analogously to the shattering case but using the properties of the small grain population:
\begin{equation}
\begin{split}
t_{\rm coa} =\;& 0.27\,\frac{a_{0.005}\,s_3}{F} \\
&\times \left(\frac{0.01}{D_{\rm S}}\right)
\left(\frac{10^3\,{\rm H\,cm^{-3}}}{n}\right)
\left(\frac{0.1\,{\rm km\,s^{-1}}}{\sigma_{\rm D,S}}\right)
\,{\rm Myr},
\end{split}
\end{equation}
where $a_{0.005}$ is the size of the small grain normalized by $5\, \rm nm$, $s_3$ is the grain density normalized by $3\, \rm g\, cm^{-3}$, $F$ is the fudge factor, $D_{\rm S}$ is the dust-to-gas ratio of small grains, $n$ is the gas density, and $\sigma_{\rm D,S}$ is the velocity dispersion of the small grains. 
Because gas densities of order $10^{2-3}\,{\rm cm^{-3}}$, which are required for a direct calculation of the coagulation timescale, are not well resolved at $68\,{\rm pc}$, we apply a sub-grid treatment.
We assume that a fraction $F=0.5$ of the unresolved gas reaches these dense, low-dispersion conditions.

\subsubsection{Intrinsic extinction curve reconstruction}
\label{sec:subsub-intrinsic_extinction}

Following \citet{Dubois24}, we reconstruct continuous intrinsic extinction curves from the two-size dust abundances by assigning a modified log-normal size distribution per hydrogen nucleus to each grain species $i$ and size bin $j$, ${\rm d}n_{i,j}/{\rm d}a = C_{i,j}a^{-4}\exp[-\ln^{2}(a/a_j)/(2\sigma^{2})]$.
We adopt $\sigma=0.75$ and characteristic radii $a_j=0.005$ and $0.1\,\mu{\rm m}$ for the small and large bins, respectively.
For each galaxy, the normalization $C_{i,j}$ is determined from the corresponding component dust-to-gas ratio $D_{i,j}$ by requiring $\mu_{\rm H}m_{\rm H}D_{i,j} = \int (4\pi/3)s_i a^3({\rm d}n_{i,j}/{\rm d}a)\,{\rm d}a$, where $\mu_{\rm H}\equiv\rho/(m_{\rm H}n_{\rm H})=1/X_{\rm H}$ is the gas mass per hydrogen nucleus, $X_{\rm H}$ is the hydrogen mass fraction, and $s_i$ is the grain material density.

We approximate the carbonaceous dust grain as graphite and adopt smoothed-UV astronomical silicate for the silicate component.
For spherical grains, we calculate the extinction efficiency $Q_{{\rm ext},i}(a,\lambda)$ directly using Mie theory \citep{Bohren&Huffman83}, using complex refractive indices derived from the graphite and smoothed-UV astronomical-silicate dielectric functions of \citet{Draine&Lee84} and \citet{Laor&Draine93}, where $Q_{{\rm ext},i}=Q_{{\rm abs},i}+Q_{{\rm sca},i}$.
The contribution of each grain species $i$ and size bin $j$ to the extinction per hydrogen nucleus is then calculated as $A_{\lambda,i,j}/N_{\rm H} = 2.5\log_{10}(e) \int\pi a^2 Q_{{\rm ext},i}(a,\lambda) ({\rm d}n_{i,j}/{\rm d}a)\,{\rm d}a$.

The carbonaceous-small, carbonaceous-large, silicate-small, and silicate-large contributions are summed to obtain the total intrinsic extinction.
The total curve and all component curves are normalized by the same total extinction in the $V$ band, evaluated at $\lambda_V=0.55\,\mu{\rm m}$.
These curves therefore describe the intrinsic extinction implied by the adopted grain mixture and do not include the effects of star--dust geometry or radiative transfer.

\section{Results and Discussion}
\subsection{Sample Galaxies}

Figure~\ref{fig:Figure1} presents mock JWST/NIRCam composite images of the sample galaxies at $z=5$, constructed from the F070W, F115W, and F200W filters without PSF convolution, using the radiative transfer code SKIRT \citep{Camps&Baes20}. 
These images are used primarily as qualitative references for the visual morphologies of the simulated galaxies, rather than as the basis for direct observational comparison or quantitative morphological measurements.
A more quantitative analysis of the mock JWST observables, including emergent attenuation curves and their dependence on star--dust geometry and viewing angle, will be presented in Jang \& Yu et al. (in preparation).
All of our analysis is done at redshift 5, the last snapshot of the NCso simulation.
We measure all physical quantities within $3R_{\rm 50}$ of each galaxy, where $R_{\rm 50}$ denotes the radius enclosing half of the total stellar mass.
Figures~\ref{fig:Figure2} and \ref{fig:Figure3} show the evolutionary tracks of two representative galaxies.
The first case develops a relatively settled disk after $z\sim5.5$, while the second illustrates a more disturbed system affected by a recent merger-driven starburst.
Figure~\ref{fig:Figure2} illustrates a galaxy that develops an early disk after $z\sim5.5$, at which point both the SFR and sSFR begin to decline.
Although the stellar and gaseous disks remain more turbulent than those of local galaxies, the gradual reduction of turbulence---driven by increased rotational support and a decreasing SFR---creates a relatively stable environment for dust growth.
The DTM distribution in Panels~(c) and (d) shows elevated DTM values (up to $\sim0.25$) along the disk plane, whereas the central region exhibits significantly lower values ($\sim0.05$).
This central low-DTM region, referred to as a ``DTM cavity'' by \citet{Byun25}, is caused by recent star formation in the bulge: clustered SN-II explosions efficiently destroy dust, lowering the central DTM relative to the disk.
Similar localized reductions in DTM can also be seen elsewhere in the disk where star formation is actively occurring (visible as small DTM-depleted holes in Panel-c).
Figure~\ref{fig:Figure3} shows an extreme case, in which a recent starburst triggered by a major merger at $z\sim5.25$ destroys a large fraction of the dust content within the galaxy.

\begin{figure}
\centering
\includegraphics[width=\columnwidth]{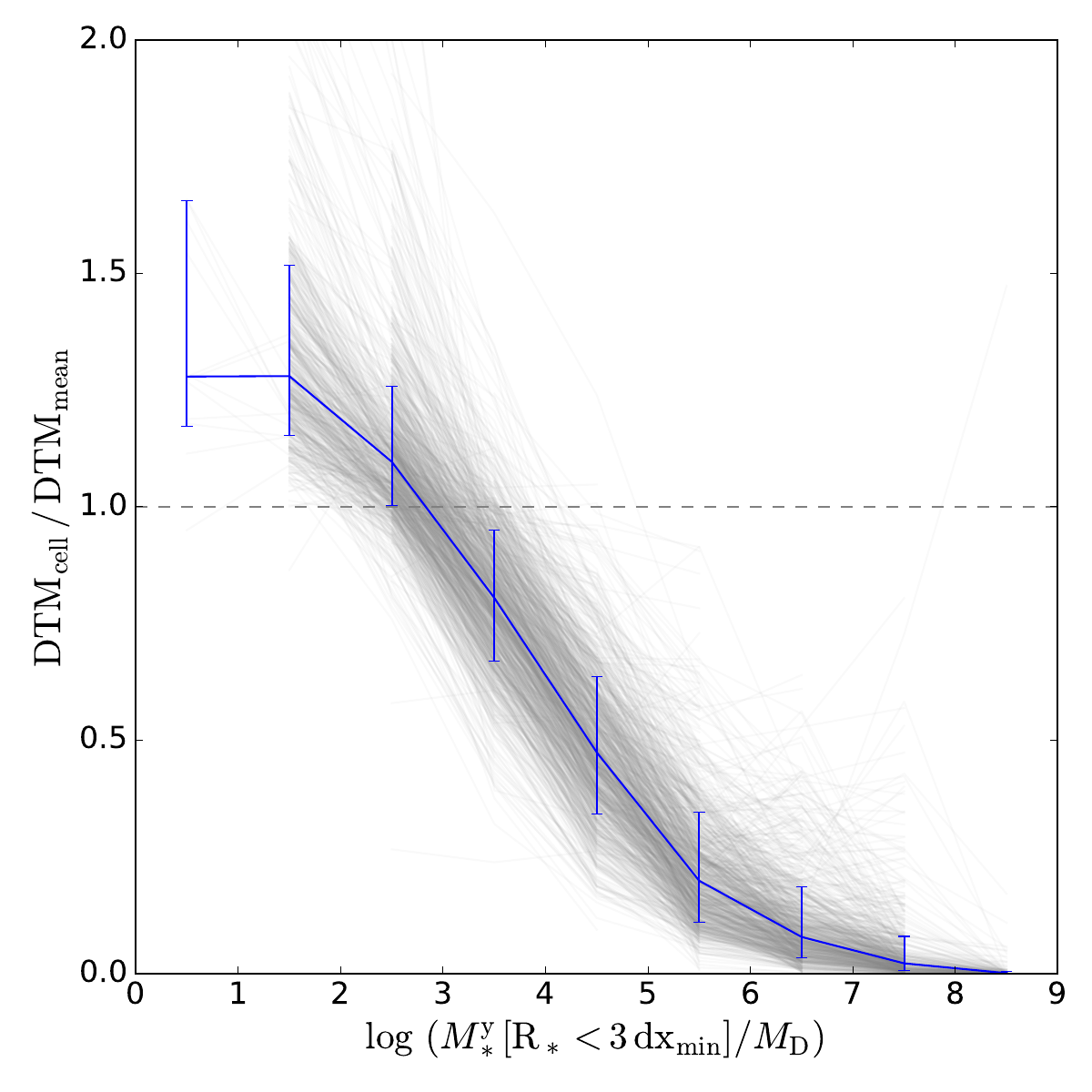}
\caption{
Dependence of the local dust-to-metal ratio (DTM) on the surrounding SN-II feedback strength.
The $x$-axis shows the relative mass ratio of young stellar mass to dust mass within $3\Delta x_{\rm min}$ ($\sim270$ pc at $z=5$) for each cell, and the y-axis shows the local DTM normalized by the mean galactic DTM.
Only young stellar particles with ages between 5 and 7.5 Myr are included, since SN-II events occur once the stellar age exceeds 5 Myr in our prescription.
Gray lines show the median trends for individual galaxies with stellar masses above $10^8\,M_\odot$.
The blue line shows the median of all gray curves, with the error bars indicating the 16th--84th percentile range.
}
\label{fig:Figure4}
\end{figure}

\begin{figure}
\centering
\includegraphics[width=\columnwidth]{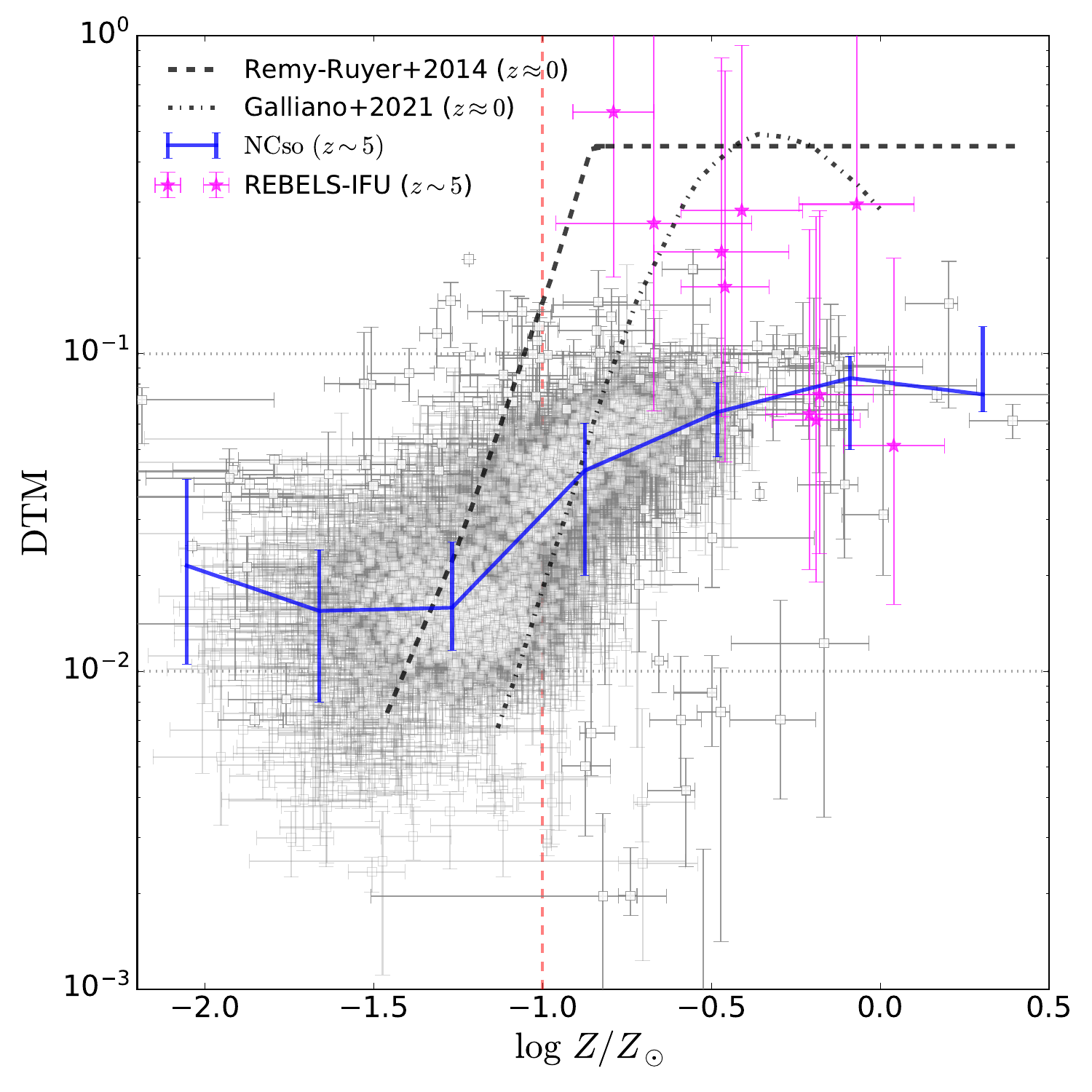}
\caption{
Dust-to-metal ratio (DTM) as a function of metallicity for galaxies at $z\sim5$.
White points represent the NCso galaxy sample. 
The error bars in both $x$ and $y$ directions show the 16th--84th percentile range of each galaxy.
The blue line and its error bars show the median DTM trend and the corresponding 16th--84th percentile range.
Magenta stars indicate the REBELS-IFU measurements from \citet{Algera26}.
The dashed and dot--dashed black lines show the DTM--metallicity relations inferred from local observations, taken from \citet{RR14} and \citet{Galliano21}, respectively.
The vertical red dashed line marks the metallicity threshold above which gas-phase metal accretion becomes the dominant dust growth channel in our simulation.
}
\label{fig:Figure5}
\end{figure}

\subsection{Dust destruction efficiency}

Figures~\ref{fig:Figure2} and \ref{fig:Figure3} clearly demonstrate that local star-formation events are closely associated with dust destruction.
To further quantify the efficiency of dust destruction on local scales, we construct a proxy that measures how susceptible the dust in each gas cell is to nearby recent star-formation activity.
For each gas cell, we compute $M_{\star}^{\rm y}$, the total mass of stellar particles with ages between 5 and 7.5 Myr within $3\Delta x_{\rm min}\simeq270\,{\rm pc}$.
We then evaluate $M_{\star}^{\rm y}/M_{\rm D}$, where $M_{\rm D}$ is the dust mass of the cell.
Figure~\ref{fig:Figure4} shows ${\rm DTM}_{\rm cell}/{\rm DTM}_{\rm mean}$ as a function of $\log(M_{\star}^{\rm y}/M_{\rm D})$, where ${\rm DTM}_{\rm cell}$ is the dust-to-metal ratio of each cell and ${\rm DTM}_{\rm mean}$ is the mass-weighted mean dust-to-metal ratio of gas cells within $3R_{\rm 50}$.
Each gray curve shows the median trend for an individual galaxy with a stellar mass greater than $10^{8}\,M_\odot$, while the blue line shows the overall median relation and the 16th--84th percentile range.
We find a consistent declining trend of local DTM with increasing values of our proxy, with relatively small scatter across the sample.
A decrease in DTM with increasing SN activity is naturally expected, but the remarkable similarity of the trends across different galaxies is noteworthy.
Although this behaviour is a direct consequence of our dust destruction prescription, it remains valuable to measure and quantify such a proxy from the final simulation output---particularly given the cosmological context and finite resolution of our simulation.
This relation may provide a useful simulation-based diagnostic for assessing how unresolved clustered feedback affects local dust survival, especially in models or post-processing calculations without explicit dust destruction.

\begin{table*}
  \centering
  \caption{Definitions and key characteristics of \textbf{Phases I--III}.}

  \begin{tabular}{ccccc}
  \hline \hline 
 Phase$^a$ 
 & $\log_{10}(Z/Z_\odot)$$^b$ 
 & Main character$^c$ 
 & Gas-phase metal accretion efficiency$^d$ 
 & Shape$^e$ \\ [3.5pt]
  \hline \hline 

 \textbf{Phase I} 
 & $< -1$ 
 & Stellar ejecta dominated
 & $\mathrm{Sil}\, $\accdown $\ \&\ $ $\mathrm{C}\, $\accdown
 & Shallow, featureless \\ [3.5pt]

 \textbf{Phase II} 
 & $[-1,-0.5)$
 & Silicate growth
 & $\mathrm{Sil}\, $\accup $\ \&\ $ $\mathrm{C}\, $\accdown
 & SMC-like \\ [3.5pt]

 \textbf{Phase III} 
 & $\ge -0.5$ 
 & Silicate \& carbon growth
 & $\mathrm{Sil}\, $\accup $\ \&\ $ $\mathrm{C}\, $\accup
 & MW-like \\ [3.5pt]

 \hline
 \end{tabular}%
 
 \label{tab:Table1}

  \begin{minipage}{\textwidth}
  \footnotesize
  \textit{Notes.}
  $^a$ Three characteristic phases of dust evolution.
  $^b$ Metallicity range defining each phase.
  $^c$ Main physical character of each phase.
  $^d$ Relative efficiency of gas-phase metal accretion onto silicate and carbonaceous grains; upward and downward arrows indicate efficient and inefficient accretion, respectively.
  $^e$ Characteristic extinction curve shape.
  \end{minipage}
 
\end{table*}

\subsection{Dust-to-Metal Ratio}

We present the dust-to-metal ratio (DTM) of galaxies as a function of gas-phase metallicity in Figure~\ref{fig:Figure5}.
Each white square with error bars shows the median DTM and metallicity of an individual galaxy, together with their 16th--84th percentile ranges.
The blue line and its error bars represent the median DTM trend and the corresponding 16th--84th percentile range across the full NCso sample.
For comparison, the dashed and dot--dashed black lines indicate the local DTM--metallicity relations from \citet{RR14} and \citet{Galliano21}, respectively.
High-redshift observational measurements from the REBELS-IFU survey \citep{Algera26} are shown as magenta stars with error bars.
The fact that the NCso sequence lies below local DTM--metallicity relations is not unexpected, since high-redshift galaxies are gas rich, dynamically young, and have had less time for dust growth. 
A similarly low-DTM sequence is also found for isolated gas-rich disk galaxies in \citet{Dubois24}.
However, within the current comparison, our simulated galaxies do not reach the high-DTM end of the REBELS-IFU measurements, where several galaxies have median DTM values of $\gtrsim 0.2$ at $-1 \lesssim \log(Z/Z_\odot) \lesssim -0.25$.

Although the large uncertainties in individual REBELS-IFU measurements prevent a strict one-to-one assessment, it is still notable that several REBELS-IFU galaxies occupy a higher-DTM regime, in some cases approaching the local relations.
The median DTM trend of our sample shows a pronounced rise from $\sim 0.02$ to significantly higher values around ${\rm log}(Z/Z_\odot)\sim -1$.
We identify this upturn as the point where efficient gas-phase metal accretion becomes dominant once the median metallicity exceeds $\sim0.1\,Z_\odot$.
This ``threshold'' metallicity, $Z_{\rm thr}$, acting as a critical value for rapid dust growth through accretion, is consistent with findings from previous studies \citep{Li19,Hou19,Parente22}.
Despite large galaxy-to-galaxy scatter---arising from variations in star formation and merger histories---the median DTM still increases gradually with metallicity.
Interestingly, some galaxies show a decrease in DTM at ${\rm log}(Z/Z_\odot)\geq 0$, a feature that also appears in the sample median.
Some REBELS-IFU galaxies also occupy a similar high-metallicity, low-DTM region, although the physical origin of those systems cannot be uniquely inferred from the present comparison. 
In NCso, the corresponding outliers are galaxies that have recently experienced merger-driven starbursts, where clustered SN-II feedback efficiently destroys dust and lowers the DTM.

\subsection{Grain evolution}

\begin{figure}
\includegraphics[width=\columnwidth]{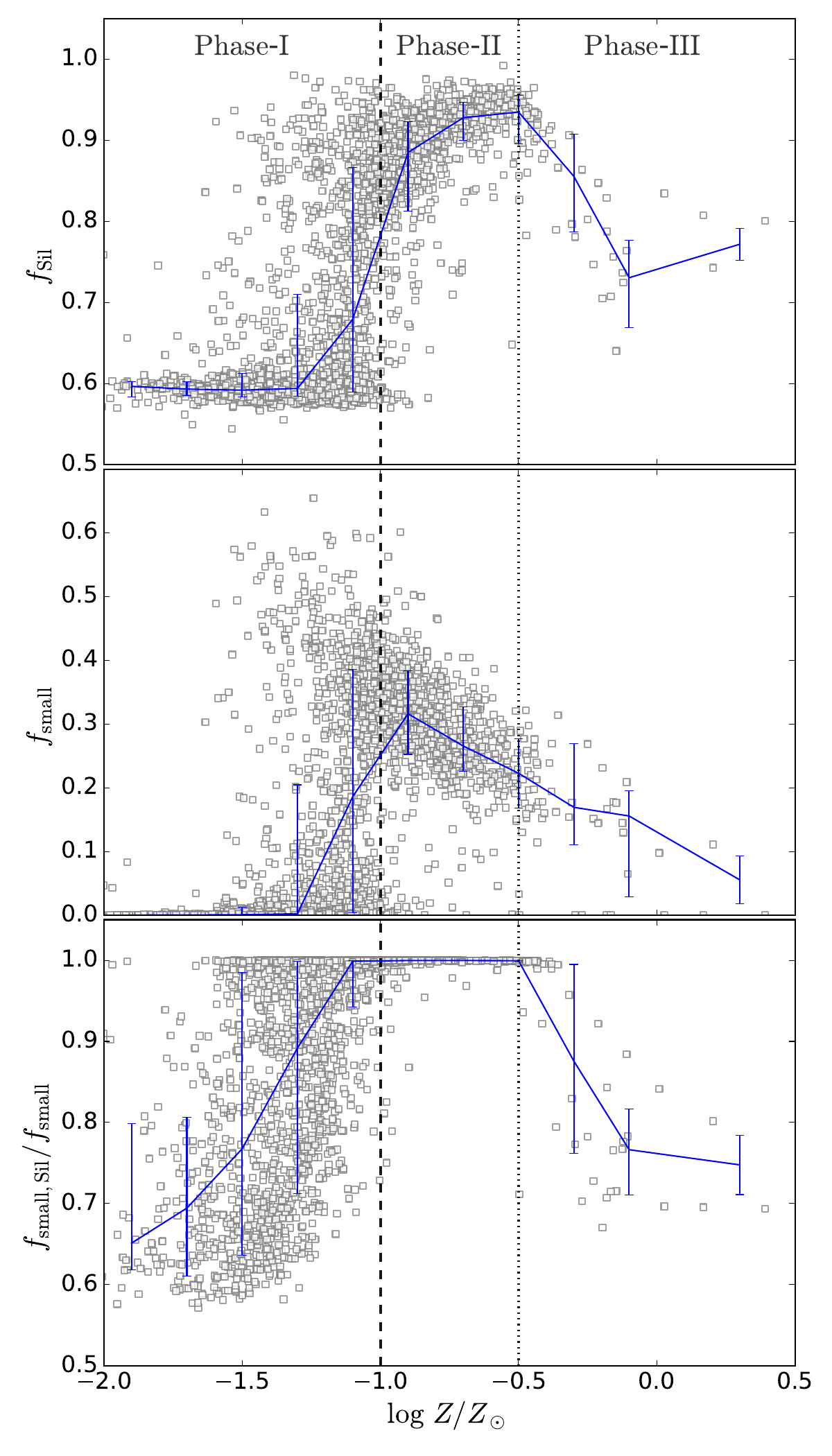}
\centering
\caption{
Grain properties as a function of gas-phase metallicity for the NCso galaxies at $z\sim5$.
White squares represent the median values for individual galaxies, and the blue line with error bars shows the sample median and the 16th--84th percentile range.
The upper, middle, and lower panels show the silicate mass fraction, the small-grain mass fraction, and the fraction of small silicate grains among all small grains, respectively.
The vertical dashed and dotted lines mark the metallicity boundaries used to divide the sample into \textbf{Phases I}, \textbf{II}, and \textbf{III}.
}
\label{fig:Figure6}
\end{figure}

We present the grain-evolution properties of our sample galaxies in Figure~\ref{fig:Figure6}.
Panel (a), (b), and (c) show, respectively, the silicate mass fraction ($f_{\rm Sil}$), the small-grain mass fraction ($f_{\rm small}$), and the fraction of small silicate grains among all small grains ($f_{\rm small,Sil}/f_{\rm small}$) as functions of gas-phase metallicity.
Two vertical demarcation lines indicate the threshold metallicities that separate three characteristic evolutionary phases: \textbf{Phase I, II, and III}.
We find that the dust properties of our simulated galaxies follow a metallicity-dependent evolutionary sequence that can be divided into three characteristic phases.
The physical interpretation of this sequence should be considered together with several explicit choices in our dust model. 
Stellar ejecta inject only large grains by construction, carbon accretion is suppressed above the adopted CO lock-up threshold, and the fixed Coulomb factors preferentially enhance accretion onto small silicate grains in cold, dense gas. 
These choices contribute directly to the characteristic grain-size and composition contrasts among the three phases. 
However, the metallicities at which the transitions occur, their relative prominence, and the galaxy-to-galaxy scatter are not imposed a priori;
they arise from the interaction of these prescriptions with the simulated chemical-enrichment histories, ISM conditions, and dust processing.

\begin{figure}
\includegraphics[width=\columnwidth]{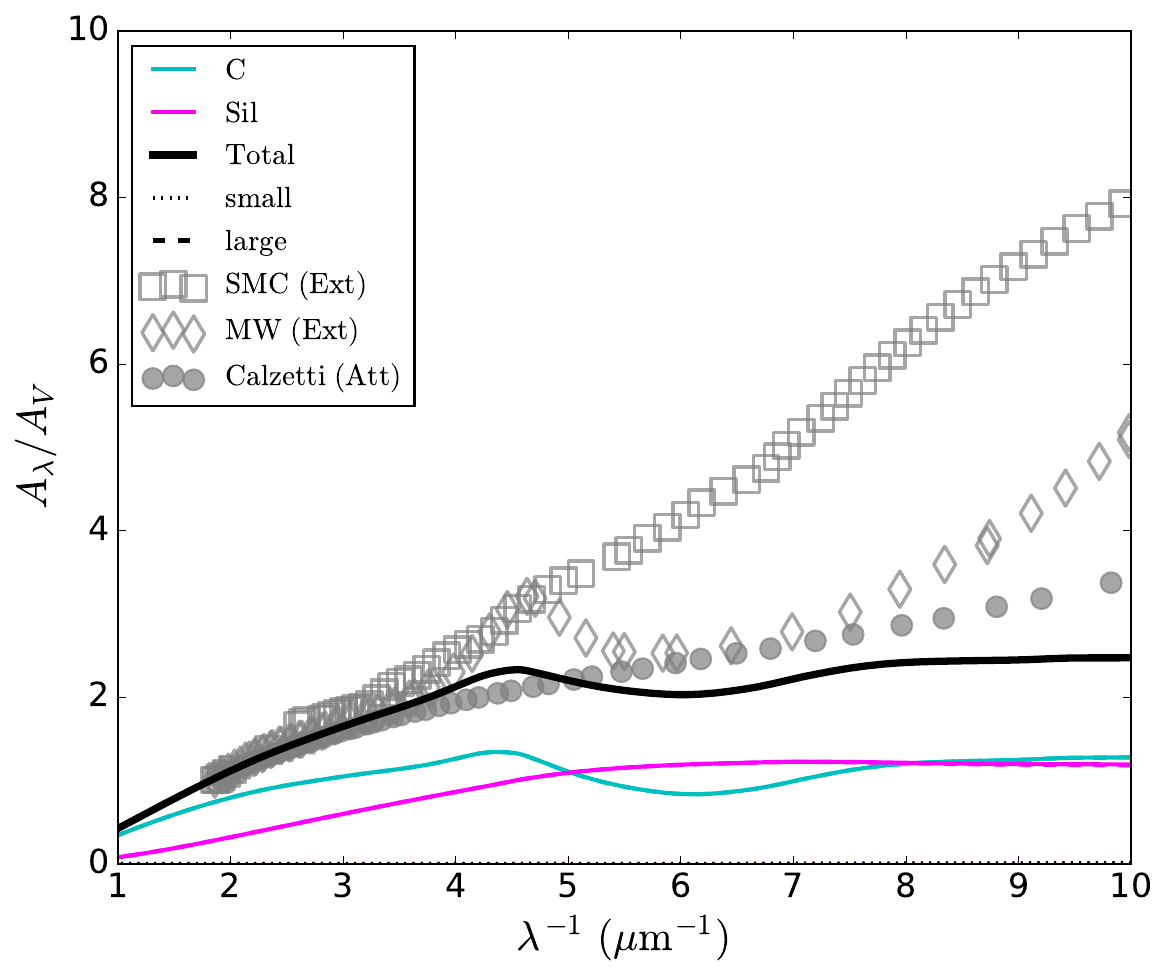}
\centering
\caption{
Representative intrinsic extinction curves for \textbf{Phase I} galaxies selected from Figure~\ref{fig:Figure6}.
The black line shows the mean total extinction curve of the selected NCso galaxies.
The magenta and cyan lines show the contributions from silicate and carbonaceous grains, respectively.
Dashed and dotted lines indicate the contributions from large and small grains, respectively.
Open gray squares and diamonds show the SMC and MW extinction curves from \citet{Pei92}, respectively, while filled gray circles show the Calzetti attenuation relation \citep{Calzetti00}.
The Calzetti relation is included only as a reference for the normalized wavelength dependence and is not treated as an intrinsic extinction curve.
}
\label{fig:Figure7}
\end{figure}

\begin{figure}
\includegraphics[width=\columnwidth]{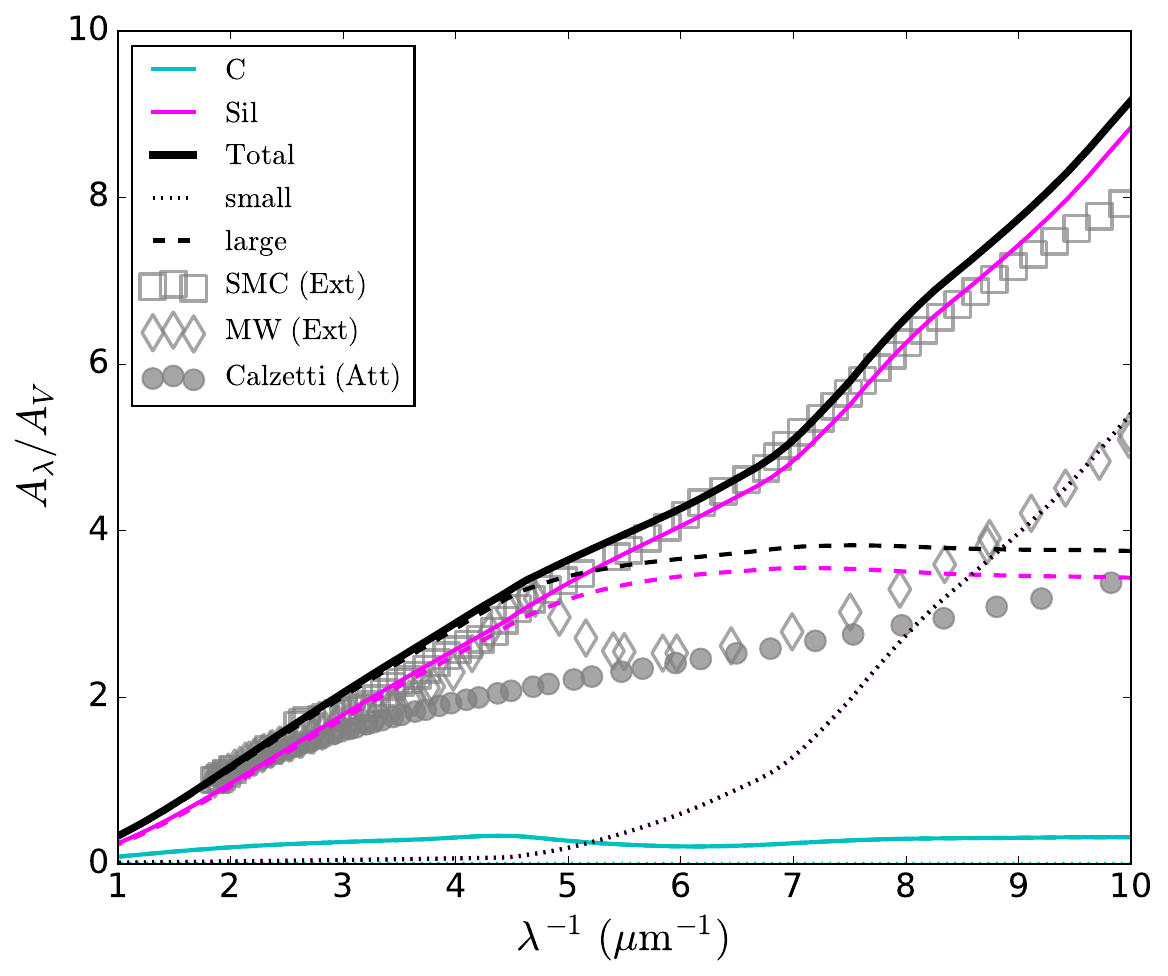}
\centering
\caption{
Representative intrinsic extinction curves for \textbf{Phase II} galaxies selected from Figure~\ref{fig:Figure6}, shown in the same format as Figure~\ref{fig:Figure7}.
The curves are SMC-like, with steep UV slopes and no prominent 2175~\AA\ feature, reflecting the dominance of small silicate grains.
}
\label{fig:Figure8}
\end{figure}

\textbf{Phase I} (${\rm log}\ Z/Z_\odot < -1$) is a {\em stellar ejecta-dominated phase}.
In this low metallicity range, gas-phase metal accretion onto pre-existing grain cores is inefficient.
The silicate fraction remains close to the intrinsic stellar yield value ($f_{\rm Sil}\approx0.6$), and the small grain fraction is nearly zero because stellar ejecta in our prescription produce only large grains.
Although shattering can generate small grains, they are quickly destroyed by thermal sputtering and SN-driven shocks.
The representative intrinsic extinction curve for this phase, derived using the methodology described in Section~\ref{sec:subsub-intrinsic_extinction} and shown in Figure~\ref{fig:Figure7}, has a shallow, featureless UV shape whose normalized wavelength dependence is similar to that of the Calzetti attenuation curve \citep{Calzetti00}.
    
\textbf{Phase II} ($-1 \leq{\rm log}\ Z/Z_\odot<-0.5$) is a {\em silicate-dominated growth phase}.
In this metallicity range, gas-phase metal accretion becomes vigorous, but it proceeds much more efficiently for silicate grains than for carbonaceous grains.
This preferential silicate growth results from the combined effects of stronger Coulomb enhancement for small silicate grains, suppressed carbon accretion in dense gas owing to CO formation, as implemented in our carbon accretion prescription, and the delayed release of carbon from AGB stars compared to the prompt SN-II enrichment of Si, Mg, and O.
Consequently, the dust population becomes strongly silicate-rich, with $f_{\rm Sil}$ exceeding 0.9 once the metallicity surpasses $\sim0.1\,Z_\odot$.
Coagulation is not yet efficient enough to convert the continuously formed small grains into larger ones, so the small grain fraction remains high.
The resulting extinction curve (Figure~\ref{fig:Figure8}) resembles the Small Magellanic Cloud (SMC) extinction curve, with a steep UV slope and no pronounced 2175~\AA\ feature.

\begin{figure}
\includegraphics[width=\columnwidth]{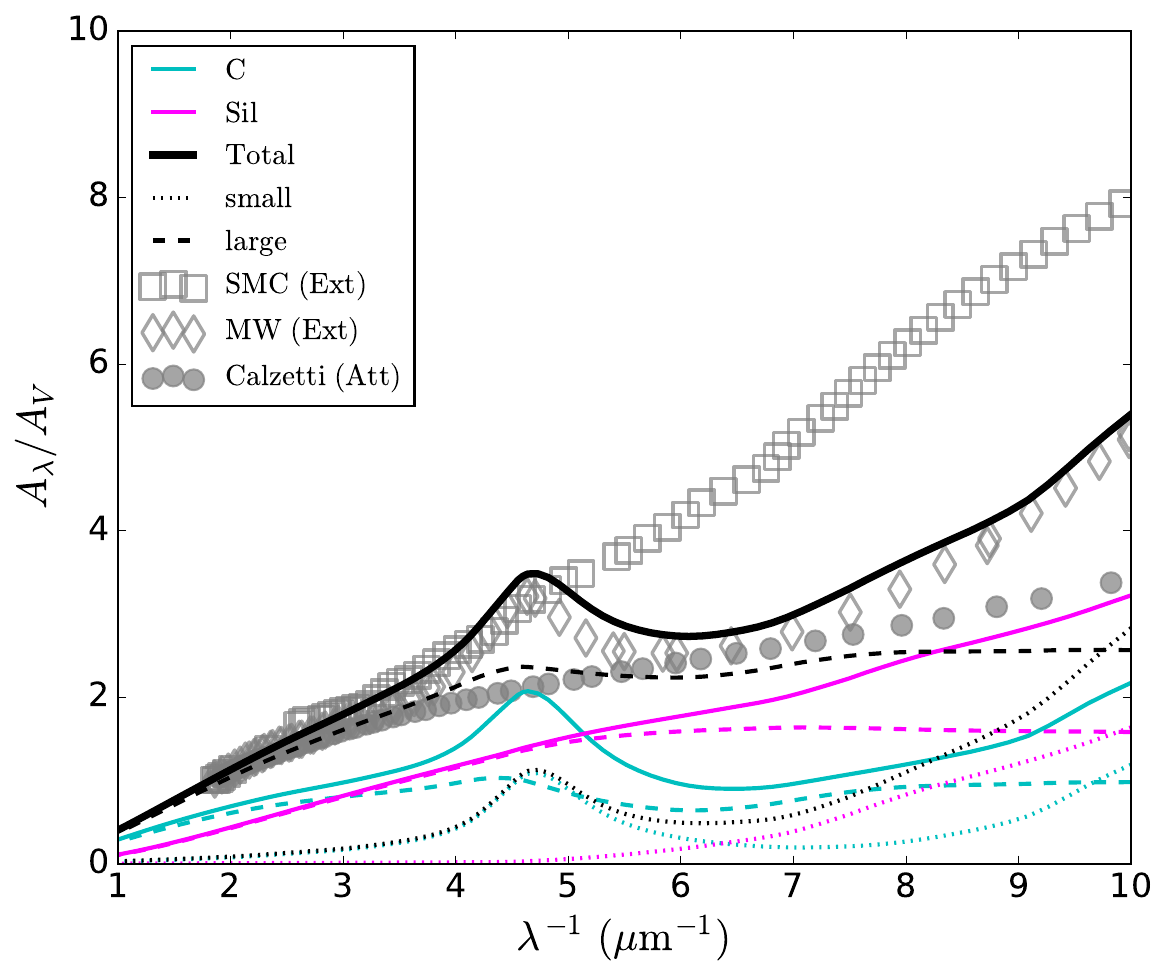}
\centering
\caption{
Representative intrinsic extinction curves for \textbf{Phase III} galaxies selected from Figure~\ref{fig:Figure6}, shown in the same format as Figure~\ref{fig:Figure7}.
The curves become more Milky-Way-like and show a noticeable 2175~\AA\ bump owing to the increased contribution from small carbonaceous grains.
}
\label{fig:Figure9}
\end{figure}

\textbf{Phase III} ($-0.5 \leq{\rm log}\ Z/Z_\odot$) marks the onset of {\em efficient accretion onto carbonaceous grains}.
At this stage, the metallicity is high enough for gas-phase accretion to proceed efficiently, while the delayed carbon enrichment from AGB stars has had sufficient time to increase the carbon budget available for dust growth.
As a result, carbon accretion becomes competitive with silicate accretion.
The silicate fraction therefore decreases to $\sim0.7$, and a more balanced small-to-large grain ratio emerges, as shown in Panel (b).
Small carbonaceous grains are also produced in non-negligible amounts (Panel c), leading to a more complex mixture of silicate and carbonaceous grains.
The corresponding intrinsic extinction curve (Figure~\ref{fig:Figure9}) becomes more Milky-Way-like and exhibits a noticeable 2175~\AA\ bump, mainly imprinted by small carbonaceous grains.

A brief summary of the definitions and key characteristics of the three phases is shown in Table~\ref{tab:Table1}.

\subsection{Dust to Stellar Mass}

\begin{figure}
\centering
\includegraphics[width=\columnwidth]{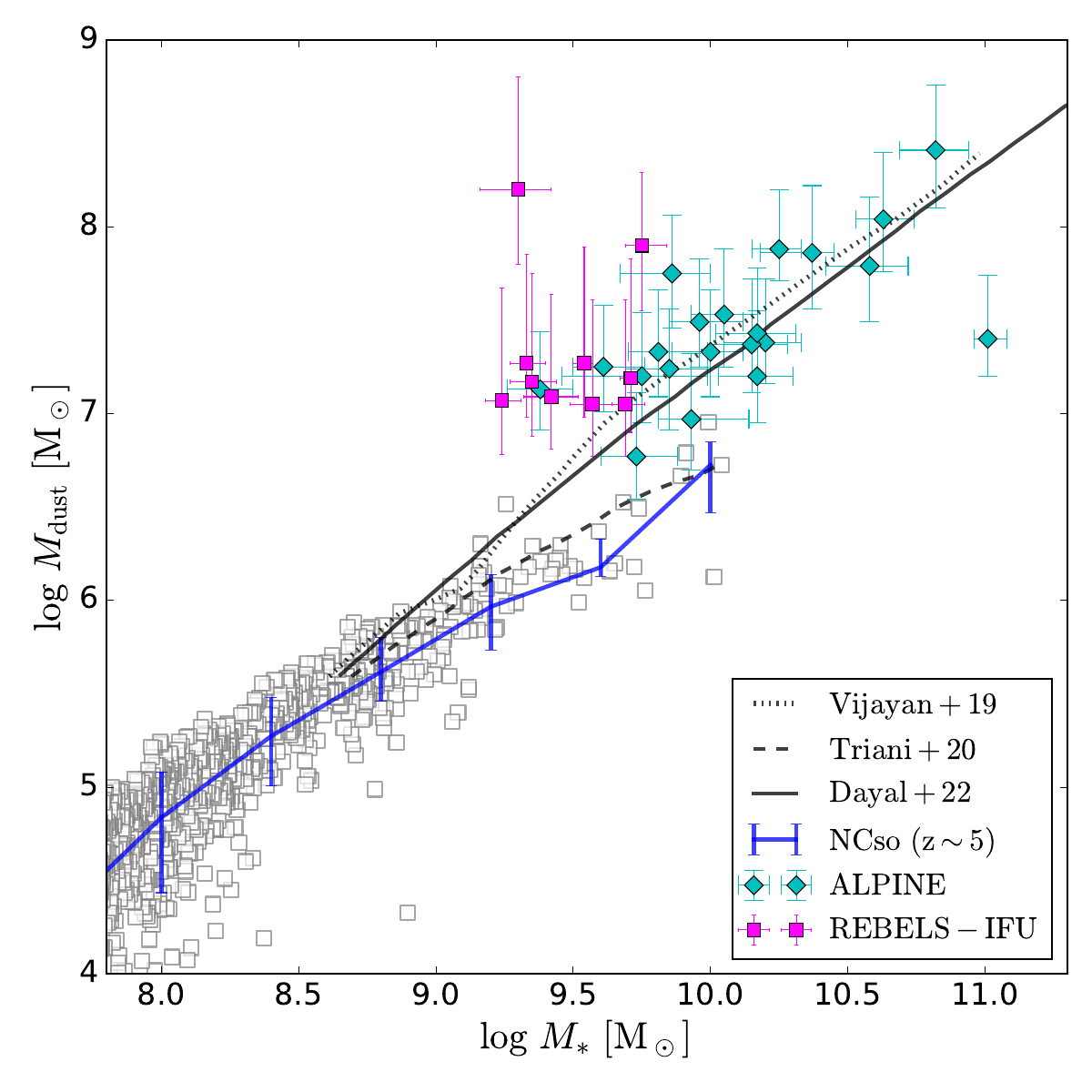}
\caption{
Relation between dust mass and stellar mass for the NCso galaxies.
White squares represent the NCso galaxy sample, and the blue line shows the median trend with the 16th--84th percentile range.
The dotted, dashed, and solid black lines indicate theoretical predictions from the semi-analytic models of \citet{Vijayan19}, \citet{Triani20}, and \citet{Dayal22}, respectively.
Magenta squares and cyan diamonds show observationally inferred dust masses from the REBELS-IFU \citep{Algera26} and ALPINE \citep{Pozzi21,Sommovigo22} surveys, respectively.
}
\label{fig:Figure10}
\end{figure}

Figure~\ref{fig:Figure10} presents the dust mass as a function of stellar mass of the galaxy.
The NCso galaxies are shown as white squares, and the blue curve indicates the median trend with the 16th--84th percentile range.
For comparison, observational measurements from ALPINE \citep{Pozzi21,Sommovigo22} and REBELS-IFU \citep{Algera26} surveys are over-plotted, together with predictions from three semi-analytic models (SAMs) \citep{Vijayan19,Triani20,Dayal22}.
We caution that this comparison is not performed in the full observational space, since the observed dust masses depend on assumptions about dust temperature, opacity, emissivity index, and the adopted FIR/mm SED model.
With this caveat in mind, NCso galaxies tend to have lower dust masses than the values inferred for observed systems over the overlapping stellar-mass range, particularly at $M_* \gtrsim 10^9\,M_\odot$.
Indeed, only a few of the most massive NCso galaxies reach the lower envelope of the observational data points.
Within NCso, the dust-mass--stellar-mass relation remains relatively shallow around $M_* \sim 10^{8.75}\,M_\odot$.
Most galaxies in this stellar-mass interval are in \textbf{Phase II}, the silicate-dominated regime identified in Figure~\ref{fig:Figure6}.
A more pronounced increase in dust mass appears only at the highest stellar masses, coinciding with the onset of efficient carbonaceous grain formation in \textbf{Phase III}.

The systematic offset relative to observations may reflect limitations in our dust model, but it may also arise from uncertainties in the observational dust mass estimation.
High-$z$ dust masses are often inferred from one or two ALMA bands, requiring strong assumptions about the dust temperature, opacity, and emissivity index.
Thus, the true explanation likely lies between model uncertainties and observational systematics.
One possible model-side concern is whether the simulation contains sufficient dust-bearing gas in the thermodynamic regime where gas-phase accretion can operate.
For the most massive NCso galaxy at the final snapshot ($M_\star\simeq10^{10}\,M_\odot$), shown in Figure~\ref{fig:Figure2}, we therefore examine the dust-bearing gas within $3R_{50}$.
Its dust-mass-weighted median gas temperature and hydrogen number density are $840\,{\rm K}$ and $31\,{\rm cm^{-3}}$, with corresponding 16th--84th percentile ranges of $430$--$2400\,{\rm K}$ and $7.3$--$72\,{\rm cm^{-3}}$, respectively.
These values indicate that a substantial fraction of the dust is associated with a relatively cool and moderately dense ISM phase in which gas-phase accretion can operate in our model.
In particular, the central temperature range remains below the $10^4\,{\rm K}$ limit adopted for the unresolved accretion treatment, and unresolved density fluctuations are statistically included through the sub-grid density PDF.
At the maximum spatial resolution of $68\,{\rm pc}$, the measured densities and temperatures represent cell-averaged conditions and do not describe the internal structure of molecular clouds.
Although this measurement is not a population-wide constraint, it shows that such conditions are present at the high-mass end of NCso.
Thus, while unresolved cloud-scale structure remains a source of uncertainty in the quantitative accretion rate, the dust-mass offset cannot be attributed simply to the absence of a suitable dust-bearing ISM phase in which the present accretion prescription can operate.
If the discrepancy mainly originates from our dust model, plausible causes include:
(i) stellar dust yields in the high-$z$ galaxies may be higher than those calibrated on local systems;
(ii) gas-phase metal accretion timescales may be shorter than adopted in our prescription;
(iii) the SN-driven destruction efficiency may be overestimated; or
(iv) the fixed grain composition may be too restrictive for chemically young high-$z$ systems.

The persistence of the dust-mass tension in the metal-enriched NCso galaxies is noteworthy because gas-phase metal accretion already dominates dust growth for galaxies with metallicities above $0.1\,Z_\odot$.
This does not imply that accretion is unimportant; rather, it indicates that the onset and integrated effect of accretion may depend on other aspects of the dust model, including the initial stellar dust supply.
\citet{Choban25} find a comparable high-redshift dust-mass tension in their fiducial simulations.
They show that low metallicities and bursty star formation can delay efficient accretion, while low initial dust abundances resulting from low SN II dust yields can prolong the subsequent dust build-up even after accretion becomes efficient.
Within their model, the stellar dust supply and subsequent accretion-driven growth are therefore not independent processes.
\citet{Narayanan26} identify local gas density and the grain-size distribution as the two leading factors governing dust growth in their simulations. 
They find that increasing turbulent velocities enhance shattering, shifting the grain-size distribution towards smaller grains and accelerating accretion through the increased grain surface area per unit dust mass. 
In NCso, the shattering timescale represents grain velocities through a prescribed density dependence and does not respond explicitly to the local gas velocity dispersion. 
It may therefore miss additional small-grain production if the turbulent high-redshift ISM raises grain collision velocities. 
Reducing the efficiency of SN-driven destruction would likewise increase the dust mass, although its quantitative importance is not isolated in the present analysis.
At $5\leq z\leq7$, \citet{Vijayan26} find that the bulk of the observed star-forming galaxy measurements, including the \citet{Algera26} sample, agree with the median dust-mass--stellar-mass relation predicted by COLIBRE. 
These comparisons demonstrate that the bulk observed relation can be reproduced by current cosmological dust models, while also emphasizing the importance of the conditions and prescriptions governing dust growth. 
In NCso, the depletion patterns discussed below additionally motivate a closer examination of the fixed grain-composition assumption.

While incorporating a dynamically evolving grain composition increases the complexity of the model, it is theoretically well-motivated---particularly in the high-$z$ environments where Fe-poor silicates, Mg-rich olivines, or alternative carbonaceous components may naturally arise due to the underlying chemical enrichment pathways.
Such a modification could directly affect gas--dust accretion efficiencies, destruction resistance, and extinction properties, potentially reconciling simulations with observed high-$z$ dust masses.

\begin{figure}
\includegraphics[width=\columnwidth]{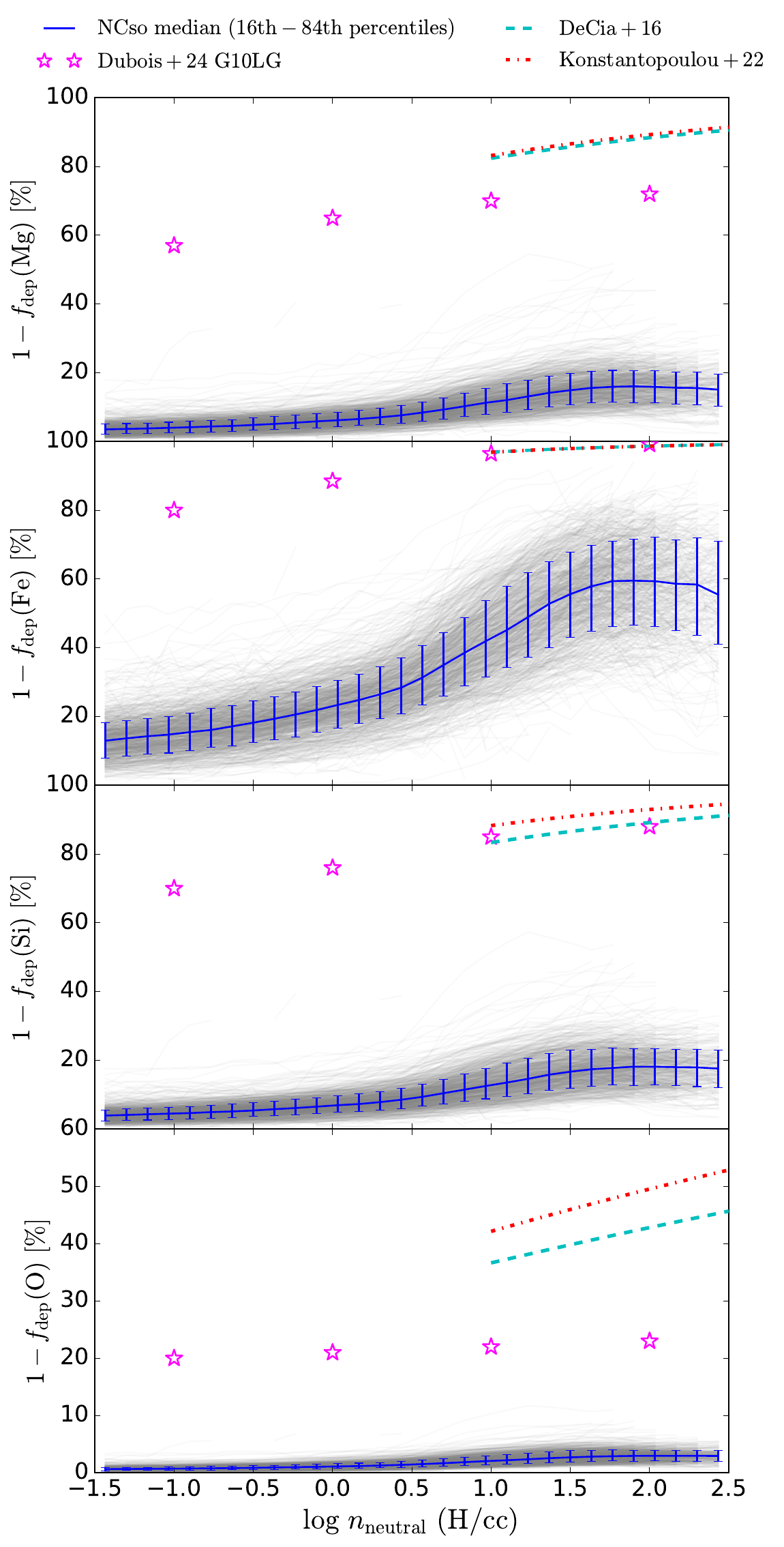}
\caption{
Median dust-phase fractions of Mg, Fe, Si, and O, expressed as $100[1-f_{\rm dep}(X)]$, as a function of the local neutral-hydrogen number density for galaxies with $M_* > 10^8\,M_\odot$.
Gray curves show the median trends of individual galaxies, while the blue curve and error bars represent the median and the 16th--84th percentile range of the galaxy population, respectively.
Magenta stars show the corresponding dust-phase fractions in the idealized isolated-disk G10LG calibration run of \citet{Dubois24}, which adopts the same dust-evolution framework but represents a more metal-rich, low-redshift MW-like disk environment.
The cyan dashed and red dash-dotted curves show the central empirical depletion sequences of \citet{DeCia16} and \citet{Konstantopoulou22}, respectively, converted to dust-phase fractions and assigned an illustrative local-density coordinate using the $[\mathrm{Zn}/\mathrm{Fe}]$--$F_*$ conversion of \citet{Richings22}, the $F_*$--$\langle n_{\rm H}\rangle$ relation of \citet{Jenkins09}, and the local-density rescaling based on \citet{Zhukovska16}. 
Because this rescaling is calibrated from Galactic Si-depletion and C\,{\sc i} constraints, and the \citet{Konstantopoulou22} sequence is parametrized by $[\mathrm{Zn}/\mathrm{Fe}]_{\rm fit}$, the observational curves are intended as reference benchmarks rather than direct element-specific measurements at fixed local density.
}
\label{fig:Figure11}
\end{figure}

To explore whether the fixed grain-composition assumption contributes to the discrepancy, we examine the depletion patterns of the elements involved in silicate formation.
Figure~\ref{fig:Figure11} presents the dust-phase fractions of the elements related to silicate grains as a function of gas density.
Here $f_{\rm dep}(X)\equiv M_{X,{\rm gas}}/M_{X,{\rm tot}}$, so that $1-f_{\rm dep}(X)$ is the fraction of element $X$ locked in dust.
Each gray curve shows the median trend for an individual galaxy, while the blue line with error bars represents the overall median and the 16th--84th percentile range.
Magenta stars indicate the corresponding dust-phase fractions for the G10LG calibration run shown in Fig.~7 of \citet{Dubois24}.
This idealized isolated-disk simulation uses the dust-evolution framework on which the NCso model is based.
The comparison is intended to illustrate the difference between the high-$z$ NCso galaxies and the calibration environment, rather than to provide a controlled one-to-one comparison.

To place the NCso element-depletion patterns in the context of empirical absorption-line measurements, we compare them with the depletion sequences of \citet{DeCia16} and \citet{Konstantopoulou22}.
These sequences are parametrized by $[\mathrm{Zn}/\mathrm{Fe}]$, which serves observationally as a tracer of depletion strength.
Since Zn is not tracked in NCso, we cannot directly evaluate these $[\mathrm{Zn}/\mathrm{Fe}]$-based sequences in our simulations.
For a qualitative comparison, we convert their empirical logarithmic gas-phase depletions, $\delta_X$, to dust-phase fractions using $1-10^{\delta_X}$.
To assign an illustrative density coordinate, we combine $F_*=1.48[\mathrm{Zn}/\mathrm{Fe}]-1.50$ \citep{Richings22} with $F_*=0.77+0.46\log_{10}(\langle n_{\rm H}\rangle/{\rm cm^{-3}})$ \citep{Jenkins09}, which gives $\log_{10}(\langle n_{\rm H}\rangle/{\rm cm^{-3}})=3.21[\mathrm{Zn}/\mathrm{Fe}]-4.93$.
We then adopt the conversion $\log_{10}(n_{\rm H,local}/{\rm cm^{-3}})=1.05\log_{10}(\langle n_{\rm H}\rangle/{\rm cm^{-3}})+2.17$, following the relation based on \citet{Zhukovska16} as written by \citet{Choban22}.
The combined mapping is therefore $\log_{10}(n_{\rm H,local}/{\rm cm^{-3}})=3.37[\mathrm{Zn}/\mathrm{Fe}]-3.01$.
Over the density range for which this mapping is applied, the central empirical dust-phase-fraction sequences of Mg, Fe, Si, and O lie above the NCso median relations.
The \citet{Konstantopoulou22} sequences are expressed in terms of their fitted $[\mathrm{Zn}/\mathrm{Fe}]_{\rm fit}$ depletion-strength estimator; treating it as a common $[\mathrm{Zn}/\mathrm{Fe}]$ proxy here is therefore an additional approximation.
Moreover, because the final density rescaling is calibrated from Galactic Si-depletion and C\,{\sc i} constraints, its application to Mg, Fe, and O should be regarded only as a common illustrative horizontal mapping rather than as an element-specific measurement of the local density.
This construction does not apply the separate $0.2$ dex adjustment to Mg, Si, and Fe adopted by \citet{Choban22} for their Milky Way comparison, because that adjustment was introduced to account for warm-neutral-medium contamination in their specific sight-line calibration and is not directly transferable to the heterogeneous samples of \citet{DeCia16} and \citet{Konstantopoulou22}.
The comparison should be interpreted as a reference benchmark rather than as a direct measurement at fixed local density, because the observed depletion sequences are derived from sight-line-integrated absorption columns, whereas the NCso relations are gas-mass-weighted profiles binned by the local gas density.
The horizontal placement also inherits the intrinsic scatter and calibration range of the empirical conversions.

We find that most of the Mg, Si, and O remain predominantly in the gas-phase rather than in the form of solid grain at $z\sim5$.
Because Fe is relatively scarce in young, SN-II-dominated systems with limited SN-Ia enrichment, the fixed Fe-bearing olivine stoichiometry makes Fe the limiting element for silicate growth.
As a result, the incorporation of Mg and Si into silicate grains becomes tied to the available Fe budget, even when Mg and Si themselves remain relatively abundant in the gas phase.
This behaviour raises the possibility that the adopted single olivine composition, $\rm Mg_{2-x}Fe_xSiO_4$ with $x=1$, may be too restrictive for chemically young high-redshift galaxies.
While this composition may successfully reproduce the bulk dust properties of local, MW-like disk galaxies, its fixed Fe-bearing olivine stoichiometry may not be directly transferable to chemically young high-$z$ systems with different Mg/Si/Fe abundance patterns.
This does not demonstrate that the Fe-rich olivine assumption is the dominant cause of the dust-mass offset.
Rather, it suggests that the present model may couple silicate growth too rigidly to a fixed Mg--Fe stoichiometry, whereas the effective Mg/Fe ratio of silicate grains could vary with the underlying chemical enrichment history.

This limitation is not restricted to the gas-phase accretion stage.
The same fixed composition assumption also enters at the stage of stellar dust injection: in our model, elemental yields from stellar ejecta are converted into silicate dust using the adopted $\rm MgFeSiO_4$ composition, together with the condensation efficiencies and the limiting-element criterion.
Thus, the model effectively assumes that the silicate dust initially supplied by SNe and AGB stars already follows a MW-like Fe-bearing olivine stoichiometry.
While this is a practical simplification for a cosmological dust model, it may be inappropriate for chemically young high-redshift systems, where the abundance pattern is dominated by prompt SN-II enrichment and has not yet reached the more Fe-enriched conditions characteristic of local MW-like galaxies.
Indeed, detailed SN dust-formation calculations generally do not predict the direct production of large amounts of Fe-bearing olivine in the ejecta.
Instead, they tend to form Mg-rich silicates, such as forsterite- or enstatite-like species, together with separate Fe-bearing or metallic grains, including Fe, Fe$_3$O$_4$, and FeS \citep{Nozawa03,Sarangi&Cherchneff15,Marassi19}.
This suggests that the fixed Fe-bearing olivine assumption may affect both the initial stellar dust supply and the subsequent ISM growth of silicate grains.

Taken together, these results suggest that the dust-mass offset may not be fully resolved by simply adjusting a single parameter such as the stellar yield, accretion timescale, or destruction efficiency.
Instead, they point to the need to examine structural assumptions in the model, especially the use of a fixed grain composition for both the initial stellar dust injection and the subsequent ISM growth.
A dust model that allows metallicity- or time-dependent variations in silicate and carbonaceous components, including Mg-rich or Fe-poor silicate pathways, may therefore provide a promising direction for future work.
Developing such a framework will demand both improved theoretical constraints on high-$z$ dust mineralogy and dedicated observational efforts to probe the elemental depletion patterns and extinction properties of early galaxies.
This work highlights the importance of moving beyond fixed grain compositions when modelling dust evolution in the early Universe, and provides a physically motivated direction for future refinements of cosmological dust prescriptions.

\subsection{Caveats and Future Directions}
Although our analysis highlights several possible directions for improving high-redshift dust modelling, a number of caveats must be acknowledged.

First, as we discussed in the previous section, the dust prescription implemented in NCso over-simplifies the underlying grain mineralogy.
The strong Fe depletion seen in Figure~\ref{fig:Figure11} suggests that the assumed Fe-bearing olivine stoichiometry may not fully capture the silicate formation pathways in chemically young galaxies.
Possible alternatives include Mg-rich olivines, pyroxene-like species, or separate nano-scale Fe-bearing components, which would change how the available elements are partitioned among dust species and could alter the resulting accretion, depletion, and extinction properties.

The sensitivity to grain chemistry is illustrated by other simulation models. 
\citet{Choban22} show that their fixed-composition Species model broadly reproduces the observed Mg and C depletion trends, while the agreement for Si depends on the adopted stellar yields.
Additional O-reservoir and nano-iron components are required to reproduce the observed O and Fe depletion trends.
For comparison, COLIBRE follows graphite together with the Mg- and Fe-endmember olivines forsterite and fayalite, each in small and large size bins, and allows the relative abundance of the two silicate species to vary with the local elemental abundances \citep{Trayford26}.

Second, the adopted two-size approximation, while computationally efficient and able to capture the bulk behaviour of grain-size evolution, cannot fully describe non-linear size-dependent processes such as charge-dependent accretion and shattering cascades.
Recent molecular-dynamics calculations of amorphous silicate grains with radii of $5$--$50$\,\AA\ ($0.5$--$5\,\mathrm{nm}$) find shattering thresholds of $\sim6\,{\rm km\,s^{-1}}$ and shattered fractions and fragment-size distributions that are not reproduced by commonly used analytic prescriptions \citep{Esmerian26a}.
Although these calculations do not extend to the $0.1\,\mu{\rm m}$ large-grain bin adopted in NCso, they demonstrate that the collision outcomes underlying simplified shattering prescriptions remain uncertain.
Although our results show clear metallicity-dependent phase transitions in grain populations, a more finely sampled grain size distribution may reveal additional pathways for small-grain regeneration or survival, particularly in merger-driven and bursty systems.
Inclusion of the multi-bin size distribution would allow more realistic modelling of the physical processes.

Third, the sub-grid treatment of gas-phase metal accretion relies on a turbulence-driven density PDF and a simplified implementation of the Coulomb enhancement factors.
\citet{Esmerian26b} find sticking coefficients of $\gtrsim0.2$ for non-inert elements on amorphous-carbon surfaces over the gas and grain temperatures considered in their calculations.
Such element- and material-specific collision physics is not represented by the single effective sticking coefficient adopted in NCso, and the calculations for carbonaceous surfaces do not directly constrain sticking onto silicate grains.
Although the sub-grid accretion is based on analytic expectations, \citet{Dubois24} have already shown that the resulting dust mass is sensitive to numerical resolution.
Such limitations are expected to be even more pronounced in the high-$z$ environment, where the ISM exhibits stronger turbulence, rapid gas cycling, and a highly complex multiphase structure.
These conditions amplify the mismatch between the true small-scale grain--gas interactions and the simplified sub-grid formulation, potentially biasing both the accretion efficiency and the relative growth of different grain populations.
For the Coulomb factors, \citet{Curro26} report substantial discrepancies with the silicate values of \citet{Weingartner&Draine99} in the CNM, which they suggest arise from differences in the treatment of grain charging.
Moreover, the accretion rate in our model is computed without a live evaluation of key physical quantities---such as the local electron density, grain charge state, or magnetic field strength---which further exacerbates the uncertainty.
This issue is also relevant for carbonaceous grains, whose accretion threshold density depends sensitively on the local cosmic ray ionization rate (CRIR).
A fully self-consistent treatment would require radiation, magnetic fields, cosmic-ray transport, and substantially higher spatial resolution.
Achieving such a treatment at the mass and spatial resolution of the present cosmological zoom-in simulation would be computationally prohibitive.
Thus, practical progress requires exploring improved sub-grid prescriptions or hybrid approaches that capture the essential physics while remaining computationally feasible---identifying, in effect, a physically motivated and numerically robust ``trade-off regime'' that balances realism with tractability.

Fourth, our simulation treats dust as a passive scalar perfectly coupled to the gas.
In reality, dust--gas drift, radiation pressure, and grain charging can induce differential motion between dust and the surrounding ISM.
These effects may be particularly important at high redshift, where radiation fields and outflow rates are strong.
Including dust--gas decoupling may alter the amount of dust reaching dense star-forming regions or escaping in galactic winds, influencing both dust growth and survival.

Finally, observational uncertainties remain substantial.
Weak constraints on dust temperature, single-band ALMA mass estimates, and the uncertain FIR-to-mm opacity can produce systematic uncertainties approaching $\sim1$ dex in the inferred dust masses.
Future ALMA and JWST programs capable of multi-band FIR/mm SED fitting, combined with rest-UV/optical extinction constraints, will be essential for breaking these degeneracies.
Cross-calibrating extinction-derived and emission-derived dust masses will also help distinguish shortcomings in the dust model from weaknesses in the observational assumptions.

\section{Summary and Conclusions}

In this work, we explored the evolution of dust in high-redshift ($z\sim5$) galaxies using the \NC\ spin-off simulation (NCso), which implements the dust formation and growth model of \citet{Dubois24} within a fully cosmological framework.
Our primary goal was to quantify how dust abundance, grain-size distribution, and grain composition evolve in the early Universe, and to identify the physical processes that may contribute to the offsets between simulations and recent ALMA/JWST-inferred dust properties.

We identified three distinct evolutionary phases---\textbf{Phase I, II, and III}---arising from the interaction between metallicity-dependent enrichment, evolving ISM conditions, and the adopted dust prescriptions.
While their characteristic grain-size and species contrasts are influenced by explicit assumptions concerning stellar dust injection, CO lock-up, and Coulomb enhancement, the transition metallicities and galaxy-to-galaxy variations emerge from the simulated evolutionary histories.
At low metallicity (\textbf{Phase I}), dust is supplied almost entirely by stellar ejecta, and the grain population remains dominated by large grains with negligible small-grain contribution.
Once the metallicity exceeds $\sim 0.1\,Z_\odot$ (\textbf{Phase II}), gas-phase metal accretion becomes efficient, driving rapid growth of silicate grains, especially for small silicates, and producing SMC-like extinction curves.
At even higher metallicity (\textbf{Phase III}), accretion onto carbonaceous grains becomes effective, reducing the silicate fraction and producing Milky Way-like extinction curves featuring a 2175~\AA\ bump.
These transitions indicate that dust growth and grain properties in high-$z$ galaxies are highly sensitive to the evolving chemical environment.

We find that local dust destruction correlates strongly with recent SN-II activity.
Our proxy---based on the ratio of nearby young stellar mass to local dust mass---reveals a consistent decline of the local DTM with increasing SN-driven feedback across the analysed galaxy sample.
This relation may provide a useful simulation-based diagnostic for assessing how unresolved clustered feedback affects local dust survival, especially in models or post-processing calculations without explicit dust destruction.

NCso galaxies tend to lie below the dust masses and dust-to-metal ratios inferred for several observed high-$z$ galaxy samples.
While the limited timescales available for growth at high-$z$ can explain part of this offset, our simulations do not reach the highest DTM values inferred for several REBELS-IFU galaxies, particularly at metallicities between $\log(Z/Z_\odot)\approx -1$ and $-0.25$.
Our sample also under-produces dust by up to $\sim0.5$--1 dex relative to several observed high-$z$ galaxies.
While observational uncertainties---particularly dust-temperature assumptions in single-band ALMA estimates---can account for variations of up to $\sim1$ dex, our results suggest that limitations in the current dust model may also contribute to the offset.

By examining elemental depletion patterns, we find that Mg, Si, and O remain largely in the gas phase at $z\sim5$, while Fe is strongly depleted into dust, reflecting limited SN-Ia enrichment in the early Universe.
This suggests that the adopted single Fe-bearing olivine composition may be too restrictive at high redshift, where Fe-poor or Mg-rich silicate pathways---and possibly pyroxene-like or nano-Fe components---could become relevant.
Such a mismatch in grain mineralogy can influence both accretion efficiency and destruction resistance, potentially contributing to the underestimation of the total dust mass.

Across these findings, we identify several structural limitations of the current model.
First, the fixed grain composition may suppress the chemically evolving mineralogy expected at high redshift.
Second, the two-size approximation restricts non-linear, size-dependent processes such as charge-dependent accretion and shattering cascades.
Third, the sub-grid accretion prescription---based on a turbulence-driven density PDF and simplified Coulomb enhancement factors---cannot fully reproduce the complex, multiphase nature of the $z\sim5$ ISM.
These issues are further exacerbated by the absence of local radiation fields, electron densities, magnetic fields, or cosmic-ray ionization rates, all of which strongly affect grain charging and CO chemistry.
Fourth, the perfect dust--gas coupling neglects dust drift, radiation pressure, and drag-driven segregation, which may be significant in high-$z$ starburst environments.
In addition, observational uncertainties remain substantial, especially when dust masses are inferred from limited FIR/mm photometry.

Altogether, our analysis suggests that the dust-mass discrepancy is unlikely to be fully resolved by modifying a single parameter, such as the stellar dust yield, destruction efficiency, or accretion timescale.
Instead, our results motivate the exploration of a more flexible dust-evolution framework in which grain composition, size distribution, and charging properties can vary with the evolving chemical state of the galaxy.
Future progress will require two complementary directions:
(i) improved theoretical models incorporating metallicity-dependent silicate compositions, multi-bin grain-size distributions, dust--gas decoupling, and more physically motivated sub-grid accretion and destruction treatments; and
(ii) improved observational constraints, especially multi-band ALMA and JWST measurements capable of jointly constraining extinction curves, elemental depletion patterns, and dust temperatures in early galaxies.
By systematically identifying the physical and numerical limitations of the current framework, this work outlines a physically motivated path toward the next generation of cosmological dust models capable of capturing the complex and rapidly evolving dust physics of early galaxies.

\section*{Acknowledgements}
S.K.Y. acknowledges support from the Korean National Research Foundation (RS2025-00514475; RS-2022-NR070872). 
J.K.J. acknowledges support from the Yonsei post-doctoral program (2026-12-0044).
This work was partially supported by the Institut de Physique des deux infinis of Sorbonne Universit\'{e} and by the ANR grant ANR-25-CE31-4684 of the French Agence Nationale de la Recherche.
S.J. acknowledges support from the Korea Astronomy and Space Science Institute under the R\&D program (Project No. 2026-1-831-03) supervised by the Ministry of Science and ICT.
This work was granted access to the HPC resources of KISTI under the allocation KSC-2024-CHA-0009.
The large data transfer was supported by KREONET, which is managed and operated by KISTI.

\section*{Data Availability}
The data underlying this article will be shared on reasonable request to the corresponding author.


\bibliographystyle{mnras}
\bibliography{ref}

\bsp
\label{lastpage}
\end{document}